\documentclass[prd,aps,a4paper,superscriptaddress,twocolumn,nofootinbib]{revtex4}
\usepackage{graphicx}
\usepackage{color}
\usepackage{dcolumn}
\usepackage{bm}
\usepackage{slashed}
\usepackage{amsmath}
\usepackage{latexsym}
\usepackage{amssymb}
\usepackage{mathrsfs}
\usepackage{amsfonts}
\usepackage{url}
\allowdisplaybreaks
\begin{document}
\title{Upgraded waveform model of eccentric binary black hole based on effective-one-body-numerical-relativity for spin-aligned binary black holes}

\author{Xiaolin Liu}
\affiliation{Department of Astronomy, Beijing Normal University,
Beijing 100875, China}
\author{Zhoujian Cao
\footnote{corresponding author}} \email[Zhoujian Cao: ]{zjcao@amt.ac.cn}
\affiliation{Department of Astronomy, Beijing Normal University,
Beijing 100875, China}
\affiliation{School of Fundamental Physics and Mathematical Sciences, Hangzhou Institute for Advanced Study, UCAS, Hangzhou 310024, China}
\author{Lijing Shao} 
\affiliation{Kavli Institute for Astronomy and Astrophysics, Peking University, Beijing 100871, China}

\begin{abstract}
Effective one body numerical relativity waveform models for spin aligned binary black holes (SEOBNR) are based on the effective one body theoretical framework and numerical relativity simulation results. SEOBNR models have evolved through version 1 to version 4. We recently extended SEOBNRv1 model to SEOBNRE (Effective One Body Numerical Relativity waveform models for Spin aligned binary black holes along Eccentric orbit) model which is also valid for spin aligned binary black hole coalescence along eccentric orbit. In this paper we update our previous SEOBNRE model to make it consistent to SEOBNRv4 which is the most widely used SEOBNR waveform model. This upgraded SEOBNRE model improves accuracy compared to previous SEOBNRE model, especially for highly spinning black holes. For spin aligned binary black holes with mass ratio $1\leq q\lesssim10$, dimensionless spin $-0.9\lesssim\chi\lesssim0.995$ and orbital eccentricity $0\leq e_0\lesssim0.6$ at reference frequency $Mf_0=0.002$ ($M$ is the total mass of the binary black hole, $f_0\approx 40\frac{10{\rm M}_\odot}{M}$Hz), the upgraded SEOBNRE model can always fit numerical relativity waveform better than 98.2\%. For most cases the fitting factor can even be better than 99\%.
\end{abstract}

\maketitle

\section{Introduction}
Tens of compact binary coalescence (CBC) events have been detected by LIGO and VIRGO \cite{2018arXiv181112907T,Nitz_2019a,Nitz_2019b,Magee_2019,PhysRevD.100.023007,PhysRevD.100.023011}. One possible channel for the formation of merging binaries is isolated evolution in the field. Another possible channel is through dynamical interaction in dense stellar environments such as globular clusters or galactic nuclei. Currently it is not clear the detected binaries are formed through which channel. The binaries formed in the field can radiate away their eccentricity. The dynamically formed binaries may still have significant residual eccentricity when their gravitational waves enter the LIGO-Virgo band. People may infer the formation channel of the binary through the eccentricity detection. Consequently more and more attention is paid to the eccentricity detection recently \cite{PhysRevD.86.104027,PhysRevD.87.127501,PhysRevD.90.104010,PhysRevD.91.063004,PhysRevD.92.044034,PhysRevD.96.084046,PhysRevD.100.064006,Loutrel_2020,Loutrel:2016cdw,Loutrel:2017fgu,Loutrel:2018ssg,Moore:2018kvz,Loutrel:2018ydu,Gondan:2018khr,Gondan:2017wzd,Hoang:2017fvh,Gondan:2017hbp,Moore_2019,Moore:2019vjj,PhysRevD.101.101501,Nitz_2020,2020arXiv200514146L,PhysRevD.101.083015,ramos-buades2020impact}.

In order to estimate the eccentricity, accurate waveform model for eccentric binary system is needed. Most waveform models for eccentric binary are based on post-Newtonian approximation and are consequently valid only for inspiral part \cite{PhysRevD.80.084001,PhysRevD.82.024033,PhysRevD.90.084016,PhysRevD.91.084040,PhysRevD.93.064031,PhysRevD.93.124061,PhysRevD.98.044015,2020MNRAS.495..466W}. Currently waveform models covering the whole inspiral-merger-ringdown process for binary black hole (BBH) appear \cite{2022ApJ...936..172K}. One is the Eccentric, Nonspinning, Inspiral, Gaussian-process Merger Approximated waveform model (ENIGMA) \cite{PhysRevD.95.024038,PhysRevD.97.024031,2020arXiv200803313C} and the other three are based on effective-one-body (EOB) framework including the Effective-One-Body Numerical-Relativity waveform model for Spin-aligned binary black holes along Eccentric orbit (SEOBNRE) \cite{PhysRevD.96.044028,Romero-Shaw:2019itr,2020arXiv200106492R,PhysRevD.101.044049,PhysRevD.103.124053}, the extended TEOBiResumS\_SM model \cite{PhysRevD.102.024077,PhysRevD.101.101501,PhysRevD.106.064049} and SEOBNRv4EHM \cite{PhysRevD.104.024046,PhysRevD.105.044035}. Among all of these waveform models for eccentric binary, only SEOBNRE and the extended TEOBiResumS\_SM model can treat spinning black holes. Such kind of complete waveform model is important to treat parameter degeneracy \cite{2020arXiv200514146L}, especially between the black hole spin and orbit eccentricity.

SEOBNRE waveform model is based on the Effective-One-Body (EOB) theoretical framework and Numerical-Relativity (NR) simulations. Buonanno and Damour firstly proposed the original idea of effective one body method for binary black hole in general relativity \cite{PhysRevD.59.084006}. Later Buonanno, Pan and others for the first time \cite{PhysRevD.76.104049} combine the effective one body method with numerical relativity results to get effective one body numerical relativity (EOBNR) model for binary black hole coalescence. Aiming for the faithful waveform template for gravitational wave detection, SEOBNRv1 \cite{PhysRevD.86.024011}, SEOBNRv2 \cite{PhysRevD.89.061502}, SEOBNRv3 \cite{PhysRevD.95.024010} and SEOBNRv4 \cite{PhysRevD.95.044028} are consequently constructed. Alternative to the SEOBNR series, Nagar, Bernuzzi and others developed TEOBResumS models \cite{PhysRevD.98.104052}. Recently EOBNR models have also been developed to describe the waveform of binary neutron stars \cite{PhysRevD.95.104036,PhysRevD.99.044051,PhysRevD.99.044007}. Even for gravitational wave memory, EOBNR model can also work quite well \cite{Cao16}.

In \cite{PhysRevD.96.044028} we extended the SEOBNRv1 to SEOBNRE model which can also describe eccentric binary black hole coalescence \cite{PhysRevD.101.044049}. Previous studies indicate that SEOBNRE waveform model works quite well \cite{Romero-Shaw:2019itr,2020arXiv200106492R,PhysRevD.101.044049}. But as noted by the authors of \cite{PhysRevD.101.101501}, the SEOBNRv1 model is outdated now. Consequently SEOBNRv1 and SEOBNRE do not accurately cover high spins or mass ratios up to 10. Compared to SEOBNRv1, SEOBNRv4 admits much higher accuracy. SEOBNRv4 satisfies the accuracy requirement of current gravitational wave detectors and has been widely used in LIGO and Virgo data analysis. In the current work we upgrade our previous SEOBNRE model to let it be consistent to the waveform model SEOBNRv4. So for clarity, we call our previous model SEOBNREv1 and the current one SEOBNREv4.

Throughout this paper we will use units $c=G=1$. This paper is arranged as following. We describe the construction of SEOBNREv4 model in the Sec.~\ref{secII}. Significantly different points between SEOBNREv1 and SEOBNREv4 are pointed out there. Along with the construction process special attention is paid to the consistency check between SEOBNRv4 and SEOBNREv4 for quasi-circular binary systems. For 2000 binary black hole systems with mass ratio ranges in $[1,10]$, spin ranges in $(-1,1)$, the fitting factor between SEOBNRv4 and SEOBNREv4 is greater than 99.98\%. In the Sec.~\ref{secIII} we then validate SEOBNREv4 against the whole spin-aligned binary black hole simulations included in the SXS catalog \cite{SXSBBH}. Finally we give a discussion and a summary in the last section.

\section{Upgrade SEOBNREv1 model to SEOBNREv4 model}\label{secII}

\begin{figure*}
\begin{tabular}{cc}
\includegraphics[width=0.5\textwidth]{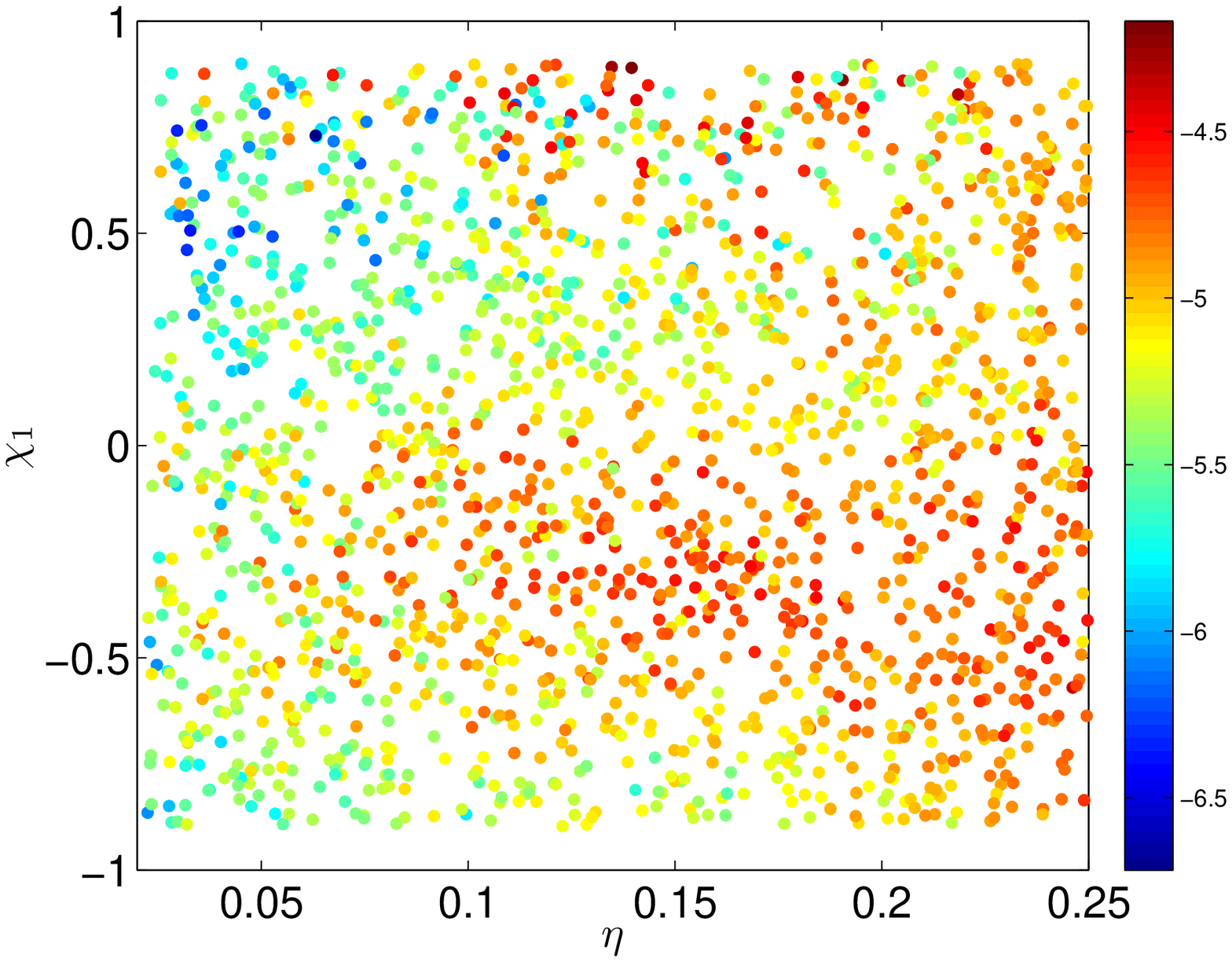}&
\includegraphics[width=0.5\textwidth]{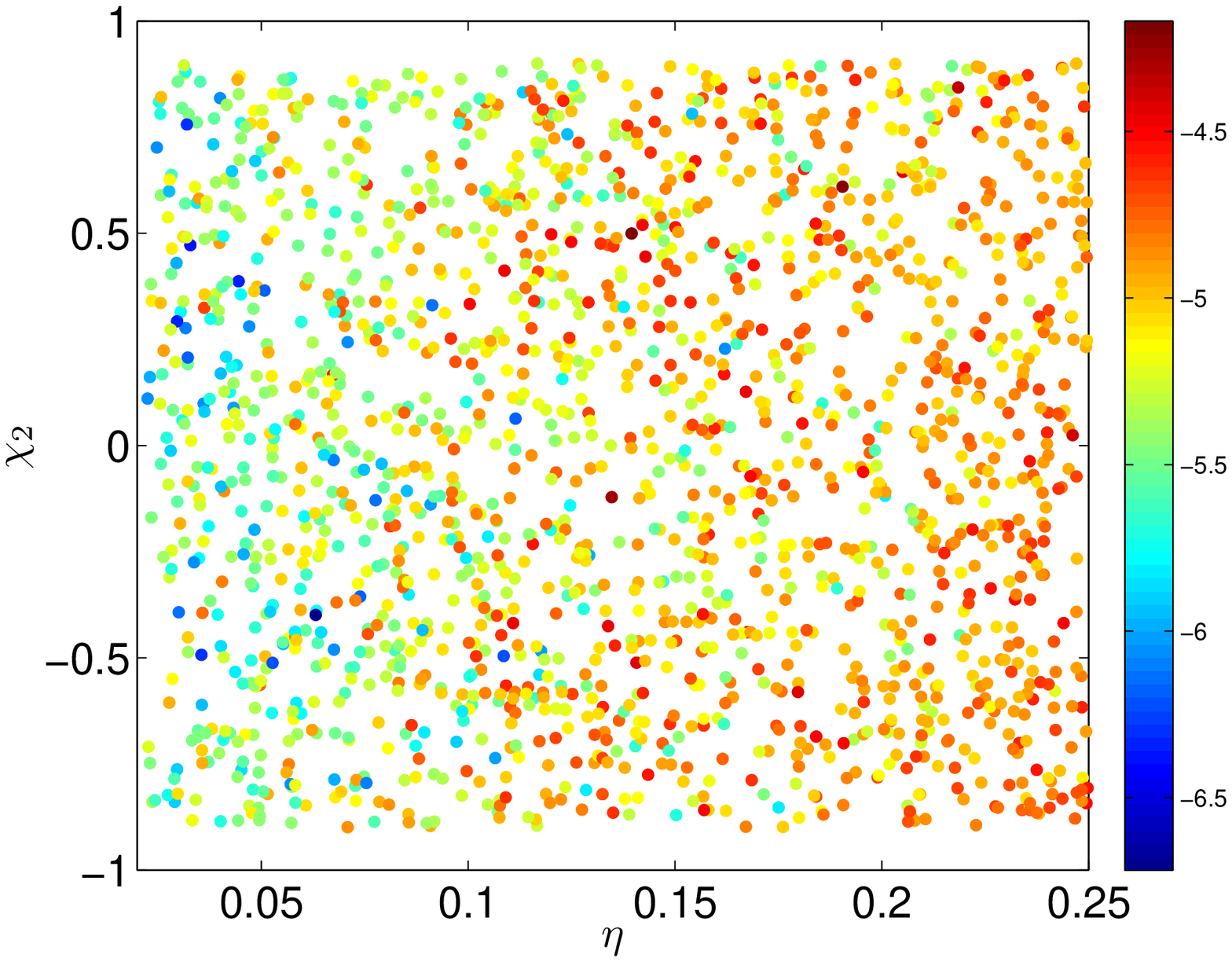}
\end{tabular}
\caption{Quasi-circular waveforms test for SEOBNREv4 against SEOBNRv4. We project the 3D parameter space of spinning, nonprecessing waveforms to the symmetric mass ratio $\eta=\frac{m_1 m_2}{(m_1+m_2)^2}$ and the two dimensionless BH spin $\chi_{1,2}$. The color represents $\log_{10}(1-{\rm FF})$.}\label{fig1}
\end{figure*}
In \cite{PhysRevD.96.044028} we constructed full inspiral-merger-ringdown waveform model for binary black hole systems along eccentric orbit based on EOBNR framework. Our construction method is independent of adiabatic approximation. So there is no direct eccentricity concept involved in our model. Instead only initial eccentricity as a parameter goes into our SEOBNRE model. This is the same as numerical relativity simulation.

Here we upgrade our previous SEOBNRE model (referred as SEOBNREv1 in the current paper for clarity) to SEOBNREv4 version which is consistent to the LIGO template SEOBNRv4 \cite{PhysRevD.95.044028}, the most widely used EOBNR waveform model so far. The basic idea of constructing SEOBNREv4 is combining the EOB dynamics with an adjusted waveform formula which describes an eccentric binary.

The SEOBNREv4 model consists two major parts, one is the waveform formula which depends on the trajectory of the effective one body; the another is the dynamics which determines the trajectory of the effective one body. Because the dynamics is divided into the conservative part and the dissipative part, the EOB dynamics also depends on the waveform formula.

Regarding to the waveform formula, we append an extra part $h^{({\rm PNE})}$ (c.f. the Eq.(52) of \cite{PhysRevD.96.044028}) to the existing waveform formula of SEOBNRv4 model \cite{PhysRevD.95.044028}. This extra waveform part describes the adjustment of eccentric motion of the binary black hole. Regarding to the dynamics part of SEOBNREv4 model, we adopt the conservative dynamics of SEOBNRv4 directly while construct the dissipative part based on the adjusted waveform according to the relations (54)-(60) of \cite{PhysRevD.96.044028}.

The nonquasi-circular correction (NQC) term $N_{\ell m}$ of the waveform formula is crucial to the quasi-circular waveform part $h^{(C)}$. The treatment of this term in SEOBNRv4 is different to that in SEOBNRv1. SEOBNRv1 assumes
\begin{align}
N_{\ell m}&=[1+\frac{\tilde{p}_r^2}{(r\Omega)^2}(a_1^{h_{\ell m}}+\frac{a_2^{h_{\ell m}}}{r}+\frac{a_3^{h_{\ell m}}+a_{3S}^{h_{\ell m}}}{r^{3/2}}\nonumber\\
&+\frac{a_4^{h_{\ell m}}}{r^2}+\frac{a_5^{h_{\ell m}}}{r^{5/2}})]\exp[i(\frac{\tilde{p}_r}{r\Omega}b_1^{h_{\ell m}}\nonumber\\
&+\frac{\tilde{p}_r^3}{r\Omega}(b_2^{h_{\ell m}}
+\frac{b_3^{h_{\ell m}}}{r^{1/2}}+\frac{b_4^{h_{\ell m}}}{r}))].
\end{align}
And the involved parameters $a_1^{h_{\ell m}}$, $a_2^{h_{\ell m}}$, $a_3^{h_{\ell m}}$, $a_{3S}^{h_{\ell m}}$, $a_4^{h_{\ell m}}$, $a_5^{h_{\ell m}}$, $b_1^{h_{\ell m}}$, $b_2^{h_{\ell m}}$, $b_3^{h_{\ell m}}$ and $b_4^{h_{\ell m}}$ have been determined as functions of binary black holes' mass ratio and spins when SEOBNRv1 was calibrated to numerical relativity. Consequently our SEOBNREv1 inherits these parameters directly.

Differently SEOBNRv4 assumes
\begin{align}
N_{\ell m}&=[1+\frac{\tilde{p}_r^2}{(r\Omega)^2}(a_1^{h_{\ell m}}+\frac{a_2^{h_{\ell m}}}{r}+\frac{a_3^{h_{\ell m}}}{r^{3/2}})]\nonumber\\
&\times\exp[i\frac{\tilde{p}_r}{r\Omega}(b_1^{h_{\ell m}}+b_2^{h_{\ell m}}\tilde{p}_r)].
\end{align}
In the implementation of SEOBNRv4, quantities $|h_{\ell m}|$, $\frac{d|h_{\ell m}|}{dt}$, $\frac{d^2|h_{\ell m}|}{dt^2}$, $\frac{d\arg(h_{\ell m})}{dt}$ and $\frac{d^2\arg(h_{\ell m})}{dt^2}$ at $t_{\rm match}$ are determined as functions of binary black holes' mass ratio and spins when SEOBNRv4 was calibrated to numerical relativity \cite{PhysRevD.95.044028}. Here the $\arg(h_{\ell m})$ means the phase of $h_{\ell m}$. When SEOBNRv4 generates a waveform for a specific binary black holes' parameters, the NQC parameters $a_1^{h_{\ell m}}$, $a_2^{h_{\ell m}}$, $a_3^{h_{\ell m}}$, $b_1^{h_{\ell m}}$ and $b_2^{h_{\ell m}}$ are solved by requiring the gotten $|h_{\ell m}|$, $\frac{d|h_{\ell m}|}{dt}$, $\frac{d^2|h_{\ell m}|}{dt^2}$, $\frac{d\arg(h_{\ell m})}{dt}$ and $\frac{d^2\arg(h_{\ell m})}{dt^2}$ at $t_{\rm match}$ equal to the ones determined by numerical relativity. We should not ask the quantities $|h_{\ell m}|$, $\frac{d|h_{\ell m}|}{dt}$, $\frac{d^2|h_{\ell m}|}{dt^2}$, $\frac{d\arg(h_{\ell m})}{dt}$ and $\frac{d^2\arg(h_{\ell m})}{dt^2}$ of the eccentric waveform equal to the ones determined through quasi-circular numerical relativity simulations. Consequently our SEOBNREv4 generates a corresponding quasi-circular waveform firstly to determine NQC parameters $a_1^{h_{\ell m}}$, $a_2^{h_{\ell m}}$, $a_3^{h_{\ell m}}$, $b_1^{h_{\ell m}}$ and $b_2^{h_{\ell m}}$. Afterwards, we use these NQC parameters to generate eccentric waveform. We caution this step is quite important. Otherwise the resulted waveform will be completely different.

For strongly eccentric orbit, the `nonquasi-circular' effect has already been counted by our $h^{({\rm PNE})}$ terms. In contrast, for quasicircular orbit, the $h^{({\rm PNE})}$ terms are negligible. Considering this fact, we introduce an adjusting factor into NQC terms. More explicitly we set following NQC for SEOBNREv4
\begin{align}
N_{\ell m}&=[1+N^{(E)}\frac{\tilde{p}_r^2}{(r\Omega)^2}(a_1^{h_{\ell m}}+\frac{a_2^{h_{\ell m}}}{r}+\frac{a_3^{h_{\ell m}}}{r^{3/2}})]\nonumber\\
&\times\exp[iN^{(E)}\frac{\tilde{p}_r}{r\Omega}(b_1^{h_{\ell m}}+b_2^{h_{\ell m}}\tilde{p}_r)],\\
N^{(E)}&=\left[1-{\rm erf}\left(\frac{Mf_0}{0.002}\frac{e_0-0.2}{0.05}\right)\right]/2.
\end{align}
Here ${\rm erf}$ means the error function. When $e_0 \rightarrow 0$, $N^{(E)}$ goes to 1 and original NQC is recovered. When $e_0$ slightly bigger than 0.2 at the reference frequency $Mf_0=0.002$, the NQC terms disappear.

About the initial state setup for the SEOBNREv4 effective one body dynamics, we firstly give some reference frequency $f_0$ of gravitational wave and the related eccentricity $e_0$ of orbit. Then we calculate the frequency according to the Kepler relation
\begin{align}
f'_0=\frac{f_0}{(1-e_0)^2}.
\end{align}
Plugging the $f'_0$ into equations
\begin{align}
&\frac{\partial H}{\partial r}=0,\\
&\frac{\partial H}{\partial p_\phi}=\frac{\pi}{f_0}
\end{align}
to solve $r_0$ and $p_{\phi_0}$ for corresponding circular orbit. Afterwards we adjust $r_0$ as
\begin{align}
r_0'=\frac{r_0}{1+e_0}
\end{align}
for warranted elliptic orbit. Then use $r'_0$ as initial setup for the SEOBNREv4 effective one body dynamics. This operation replaces the relation (67) of \cite{PhysRevD.96.044028}.

In the SEOBNRv4 code, the time evolution stops if eccentric orbit-like behavior $\dot{p}_r>0$ happens near the time when the waveform is connected to the quasi-normal modes. Although this condition implies that the test particle falls into the region inside the last stable orbit in most cases, the real orbit eccentricity (in contrast to the corrected one by the nonquasi-circular effect) possibly makes $\dot{p}_r>0$ happen outside the last stable orbit which results in waveform generation failure. Consequently we change the time evolution stop condition to requiring both the test particle falls into the region inside the last stable orbit and $\dot{p}_r>0$. This adjustment makes our SEOBNREv4 code more flexible.

As the construction of SEOBNREv1, the first check is the consistence between our SEOBNREv4 and SEOBNRv4 for quasi-circular binary black hole systems. We have done a bunch of tests for spin aligned binary black hole systems. If we plot out the two waveforms together, we can not distinguish them by eye. More quantitatively, we use the following fitting factor to describe this consistency
\begin{align}
{\rm FF}&\equiv\frac{\langle h_1|h_2\rangle}{\|h_1\|\cdot\|h_2\|},\label{equation1}\\
\langle h_1|h_2\rangle&=2\int\left(\tilde{h}_1\tilde{h}_2^*+\tilde{h}_1^*\tilde{h}_2\right)df,\\
\|h\|&\equiv\sqrt{\langle h|h\rangle},
\end{align}
where the ``$\tilde{(\cdot)}$" means the Fourier transformation, the ``${}^*$" means taking the complex conjugate. We have tested 2000 spin-aligned binary black hole systems with randomly chosen symmetric mass ratio $\eta$ in the range $[0.02,0.25]$ (corresponding to mass ratio about $1<q<50$) and the randomly chosen dimensionless spin of the black holes $\chi$ in the range $(-1,1)$. We plot the tested parameters in the Fig.~\ref{fig1}. Roughly all parameter space has been covered. For all of these tested cases, the fitting factor between SEOBNRv4 and SEOBNREv4 is bigger than 99.98\%.

In principle, Schott terms for radiation reaction force and higher multipole waveforms should be considered like recent works \cite{PhysRevD.105.044035,2022CQGra..39c5009L} for BBH along eccentric orbit. Interestingly the current work indicates that SEOBNRv4 Hamiltonian improves the waveform accuracy much better than Schott terms and higher multipoles. That is to say in EOBNR type waveform models, a good Hamiltonian is more important than Schott terms and higher multipoles for waveform accuracy.
\section{Validation of the SEOBNREv4 against numerical relativity waveforms}\label{secIII}
\begin{table*}
\centering
\caption{Validating the SEOBNRE models against the SXS small eccentricity simulations whose initial eccentricity $e_0<0.04$ when the numerical relativity simulation starts. Here ID means the SXS numbers in the catalogue \cite{SXSBBH}; $q\equiv m_1/m_2$ means the mass ratio; $\chi_{1,2}$ are the dimensionless spin parameters for the two black holes; and the fitting factors v1FF and v4FF correspond to SEOBNREv1 and SEOBNREv4 respectively. Here the fitting factor between the SEOBNRE waveforms and the numerical relativity waveforms are defined in (\ref{equation1}) with inner product (\ref{equation2}). And the values listed here are the minimal value respect to total mass between 10M${}_\odot$ and 200M${}_\odot$. `FAIL' means the waveform generation fails.}
\begin{tabular}{p{0.75cm}<\centering |p{0.8cm}<\centering |p{1.2cm}<\centering |p{1.2cm}<\centering |p{2.0cm}<\centering | p{2.0cm}<\centering || p{0.75cm}<\centering |p{0.8cm}<\centering |p{1.2cm}<\centering |p{1.2cm}<\centering | p{2.0cm}<\centering |p{2.0cm}<\centering}
\hline
ID & $q$ & $\chi_1$ & $\chi_2$ & v1FF & v4FF & ID & $q$ & $\chi_1$ & $\chi_2$ & v1FF & v4FF \\
\hline
   1 &  1.00 &   0.0000 &   0.0000 &   0.99959493 &   0.99534511 &    4 &  1.00 &  -0.5000 &   0.0000 &   0.99796292 &   0.99798073\\
   5 &  1.00 &   0.5000 &   0.0000 &   0.99808894 &   0.99921995 &    7 &  1.50 &   0.0000 &   0.0000 &   0.99967844 &   0.99448342\\
   8 &  1.50 &  -0.0000 &   0.0000 &   0.99961505 &   0.99545045 &    9 &  1.50 &   0.5000 &   0.0000 &   0.99842683 &   0.99873490\\
  12 &  1.50 &  -0.5000 &   0.0000 &   0.99901361 &   0.99751720 &   13 &  1.50 &   0.5000 &   0.0000 &   0.99880163 &   0.99905430\\
  14 &  1.50 &  -0.5000 &   0.0000 &   0.99907370 &   0.99803098 &   16 &  1.50 &  -0.5000 &   0.0000 &   0.99904231 &   0.99766303\\
  19 &  1.50 &  -0.5000 &   0.5000 &   0.99729999 &   0.99862227 &   25 &  1.50 &   0.5000 &  -0.5000 &   0.99727064 &   0.99773623\\
  30 &  3.00 &   0.0000 &   0.0000 &   0.99961316 &   0.99752402 &   31 &  3.00 &   0.5000 &   0.0000 &   0.94757016 &   0.99765069\\
  36 &  3.00 &  -0.5000 &   0.0000 &   0.99612316 &   0.99911655 &   38 &  3.00 &  -0.5000 &   0.0000 &   0.99569482 &   0.99945059\\
  39 &  3.00 &  -0.5000 &   0.0000 &   0.99551406 &   0.99954687 &   40 &  3.00 &  -0.5000 &   0.0000 &   0.99509370 &   0.99909666\\
  41 &  3.00 &   0.5000 &   0.0000 &   0.94768444 &   0.99730479 &   45 &  3.00 &   0.5000 &  -0.5000 &   0.94202563 &   0.99721943\\
  46 &  3.00 &  -0.5000 &  -0.5000 &   0.99854859 &   0.99855557 &   47 &  3.00 &   0.5000 &   0.5000 &   0.94283341 &   0.99622692\\
  54 &  5.00 &   0.0000 &   0.0000 &   0.99968717 &   0.99929932 &   55 &  5.00 &   0.0000 &   0.0000 &   0.99932697 &   0.99862830\\
  56 &  5.00 &   0.0000 &   0.0000 &   0.99943674 &   0.99861546 &   60 &  5.00 &  -0.5000 &   0.0000 &   0.99461968 &   0.99763936\\
  61 &  5.00 &   0.5000 &   0.0000 &   0.85093542 &   0.99201451 &   63 &  8.00 &   0.0000 &  -0.0000 &   0.99940263 &   0.99906153\\
  64 &  8.00 &  -0.5000 &  -0.0000 &   0.99480909 &   0.99587685 &   65 &  8.00 &   0.5000 &   0.0000 &   0.79236051 &   0.99133912\\
  66 &  1.00 &   0.0000 &   0.0000 &   0.99966398 &   0.99538394 &   70 &  1.00 &   0.0000 &   0.0000 &   0.99960646 &   0.99534181\\
  84 &  1.00 &   0.5000 &   0.0000 &   0.99768675 &   0.99916212 &   85 &  1.00 &   0.5000 &   0.0000 &   0.99759861 &   0.99893812\\
  86 &  1.00 &   0.0000 &   0.0000 &   0.99968128 &   0.99540103 &   90 &  1.00 &   0.0000 &   0.0000 &   0.99962062 &   0.99489941\\
  93 &  1.50 &   0.0000 &  -0.0000 &   0.99981259 &   0.99448156 &   93 &  1.50 &   0.0000 &  -0.0000 &   0.99981259 &   0.99448156\\
 101 &  1.50 &  -0.5000 &   0.0000 &   0.99904316 &   0.99764531 &  107 &  5.00 &   0.0000 &   0.0000 &   0.99955301 &   0.99873149\\
 109 &  5.00 &  -0.5000 &   0.0000 &   0.99500762 &   0.99812145 &  110 &  5.00 &   0.5000 &   0.0000 &   0.85695297 &   0.99169724\\
 111 &  5.00 &  -0.5000 &   0.0000 &   0.99250158 &   0.99618643 &  112 &  5.00 &   0.0000 &   0.0000 &   0.99940204 &   0.99865598\\
 113 &  5.00 &   0.0000 &   0.0000 &   0.99793174 &   0.99763316 &  114 &  8.00 &  -0.5000 &  -0.0000 &   0.99574424 &   0.99781133\\
 148 &  1.00 &  -0.4376 &  -0.4376 &   0.99806108 &   0.99424020 &  149 &  1.00 &  -0.2000 &  -0.2000 &   0.99804464 &   0.99783473\\
 150 &  1.00 &   0.2000 &   0.2000 &   0.99808965 &   0.99807992 &  151 &  1.00 &  -0.6000 &  -0.6000 &   0.99764873 &   0.99843822\\
 152 &  1.00 &   0.6000 &   0.6000 &   0.97384802 &   0.99545315 &  153 &  1.00 &   0.8500 &   0.8500 &  FAIL &   0.99621913\\
 154 &  1.00 &  -0.8000 &  -0.8000 &  FAIL &   0.99820643 &  155 &  1.00 &   0.8000 &   0.8000 &  FAIL &   0.99655259\\
 157 &  1.00 &   0.9500 &   0.9500 &  FAIL &   0.99649968 &  158 &  1.00 &   0.9700 &   0.9700 &  FAIL &   0.99400703\\
 159 &  1.00 &  -0.9000 &  -0.9000 &  FAIL &   0.99825462 &  160 &  1.00 &   0.9000 &   0.9000 &  FAIL &   0.99661560\\
 162 &  2.00 &   0.6000 &  -0.0000 &   0.95968167 &   0.99757438 &  166 &  6.00 &   0.0000 &   0.0000 &   0.99945445 &   0.99909901\\
 167 &  4.00 &   0.0000 &   0.0000 &   0.99926536 &   0.99862089 &  168 &  3.00 &   0.0000 &  -0.0000 &   0.99954667 &   0.99797457\\
 169 &  2.00 &   0.0000 &   0.0000 &   0.99966574 &   0.99648363 &  170 &  1.00 &   0.4368 &   0.4368 &   0.99607569 &   0.99269369\\
 171 &  1.00 &  -0.4379 &  -0.4379 &   0.99535381 &   0.99174050 &  172 &  1.00 &   0.9800 &   0.9800 &  FAIL &   0.99388876\\
 174 &  3.00 &   0.5000 &   0.0000 &   0.94476224 &   0.99715096 &  175 &  1.00 &   0.7500 &   0.7500 &  FAIL &   0.99662813\\
 176 &  1.00 &   0.9600 &   0.9600 &  FAIL &   0.99577598 &  177 &  1.00 &   0.9900 &   0.9900 &  FAIL &   0.99118598\\
 178 &  1.00 &   0.9950 &   0.9950 &  FAIL &   0.98839665 &  180 &  1.00 &   0.0000 &   0.0000 &   0.99975652 &   0.99556797\\
 181 &  6.00 &  -0.0000 &  -0.0000 &   0.99877816 &   0.99833008 &  182 &  4.00 &   0.0000 &  -0.0000 &   0.99912461 &   0.99811138\\
 182 &  4.00 &   0.0000 &  -0.0000 &   0.99912461 &   0.99811138 &  183 &  3.00 &  -0.0000 &  -0.0000 &   0.99952473 &   0.99701538\\
 183 &  3.00 &  -0.0000 &  -0.0000 &   0.99952473 &   0.99701538 &  184 &  2.00 &   0.0000 &   0.0000 &   0.99966817 &   0.99580905\\
 185 &  9.99 &   0.0000 &  -0.0000 &   0.99811140 &   0.99572005 &  186 &  8.27 &   0.0000 &  -0.0000 &   0.99935802 &   0.99836023\\
 187 &  5.04 &  -0.0000 &  -0.0000 &   0.99949768 &   0.99745481 &  188 &  7.19 &  -0.0000 &  -0.0000 &   0.99906217 &   0.99814348\\
 189 &  9.17 &  -0.0000 &  -0.0000 &   0.99932475 &   0.99604193 &  190 &  4.50 &  -0.0000 &  -0.0000 &   0.99966237 &   0.99744577\\
 191 &  2.51 &  -0.0000 &  -0.0000 &   0.99953024 &   0.99463385 &  192 &  6.58 &  -0.0000 &  -0.0000 &   0.99952983 &   0.99674554\\
 193 &  3.50 &  -0.0000 &  -0.0000 &   0.99960850 &   0.99677740 &  194 &  1.52 &  -0.0000 &  -0.0000 &   0.99971855 &   0.99599977\\
 195 &  7.76 &  -0.0000 &  -0.0000 &   0.99919186 &   0.99881903 &  196 &  9.66 &   0.0000 &  -0.0000 &   0.99923655 &   0.99894623\\
 197 &  5.52 &  -0.0000 &  -0.0000 &   0.99946247 &   0.99827953 &  198 &  1.20 &  -0.0000 &  -0.0000 &   0.99949745 &   0.99724950\\
 199 &  8.73 &  -0.0000 &  -0.0000 &   0.99899271 &   0.99841536 &  200 &  3.27 &  -0.0000 &  -0.0000 &   0.99926814 &   0.99520844\\
 201 &  2.32 &  -0.0000 &  -0.0000 &   0.99940447 &   0.99407207 &  202 &  7.00 &   0.6000 &  -0.0000 &   0.57728878 &   0.99224828\\
 203 &  7.00 &   0.4000 &   0.0000 &   0.92654331 &   0.99279490 &  204 &  7.00 &   0.4000 &  -0.0000 &   0.92574085 &   0.99175474\\
 205 &  7.00 &  -0.4000 &  -0.0000 &   0.99604695 &   0.99317552 &  206 &  7.00 &  -0.4000 &  -0.0000 &   0.98427562 &   0.99614997\\
 207 &  7.00 &  -0.6000 &   0.0000 &  FAIL &   0.99754335 &  208 &  5.00 &  -0.9000 &   0.0000 &  FAIL &   0.99403981\\
 209 &  1.00 &  -0.9000 &  -0.5000 &   0.99752674 &   0.99706708 &  210 &  1.00 &  -0.9000 &  -0.0000 &   0.99846264 &   0.99399869\\
 211 &  1.00 &  -0.9000 &   0.9000 &   0.99798484 &   0.99938560 &  212 &  1.00 &  -0.8000 &  -0.8000 &  FAIL &   0.99902277\\
 213 &  1.00 &  -0.8000 &   0.8000 &   0.99849551 &   0.99935168 &  214 &  1.00 &  -0.6250 &  -0.2500 &   0.99860326 &   0.99473431\\
 215 &  1.00 &  -0.6000 &  -0.6000 &   0.99843536 &   0.99834033 &  216 &  1.00 &  -0.6000 &  -0.0000 &   0.99810414 &   0.99864503\\
 217 &  1.00 &  -0.6000 &   0.6000 &   0.99910281 &   0.99820365 &  218 &  1.00 &  -0.5000 &   0.5000 &   0.99800822 &   0.99772470\\
 219 &  1.00 &  -0.5000 &   0.9000 &   0.99756967 &   0.99747917 &  220 &  1.00 &  -0.4000 &  -0.8000 &   0.99831033 &   0.99840139\\
 221 &  1.00 &  -0.4000 &   0.8000 &   0.99798314 &   0.99853768 &  222 &  1.00 &  -0.3000 &   0.0000 &   0.99701519 &   0.99630601\\
\hline
\end{tabular}\label{table1}
\end{table*}
\setcounter{table}{0}
\begin{table*}
\centering
\caption{\textit{(Continued)}}
\begin{tabular}{p{0.75cm}<\centering |p{0.8cm}<\centering |p{1.2cm}<\centering |p{1.2cm}<\centering |p{2.0cm}<\centering | p{2.0cm}<\centering || p{0.75cm}<\centering |p{0.8cm}<\centering |p{1.2cm}<\centering |p{1.2cm}<\centering | p{2.0cm}<\centering |p{2.0cm}<\centering}
\hline
ID & $q$ & $\chi_1$ & $\chi_2$ & v1FF & v4FF & ID & $q$ & $\chi_1$ & $\chi_2$ & v1FF & v4FF \\
\hline
 222 &  1.00 &  -0.3000 &   0.0000 &   0.99701519 &   0.99630601 &  223 &  1.00 &   0.3000 &   0.0000 &   0.99839706 &   0.99845059\\
 223 &  1.00 &   0.3000 &   0.0000 &   0.99839706 &   0.99845059 &  224 &  1.00 &   0.4000 &  -0.8000 &   0.99838330 &   0.99851736\\
 224 &  1.00 &   0.4000 &  -0.8000 &   0.99838330 &   0.99851736 &  225 &  1.00 &   0.4000 &   0.8000 &   0.97315379 &   0.99399307\\
 226 &  1.00 &   0.5000 &  -0.9000 &   0.99815638 &   0.99881881 &  227 &  1.00 &   0.6000 &  -0.0000 &   0.99771254 &   0.99821627\\
 228 &  1.00 &   0.6000 &   0.6000 &   0.97365710 &   0.99548934 &  229 &  1.00 &   0.6500 &   0.2500 &   0.99611746 &   0.99491419\\
 230 &  1.00 &   0.8000 &   0.8000 &  FAIL &   0.99616625 &  231 &  1.00 &   0.9000 &  -0.0000 &   0.99381213 &   0.98805670\\
 232 &  1.00 &   0.9000 &   0.5000 &   0.86425579 &   0.99610706 &  233 &  2.00 &  -0.8713 &   0.8500 &   0.98518844 &   0.99037345\\
 234 &  2.00 &  -0.8500 &  -0.8500 &  FAIL &   0.99222283 &  235 &  2.00 &  -0.6000 &  -0.6000 &   0.99843581 &   0.99930459\\
 236 &  2.00 &  -0.6000 &  -0.0000 &   0.99693930 &   0.99830705 &  237 &  2.00 &  -0.6000 &   0.6000 &   0.99127706 &   0.99600619\\
 238 &  2.00 &  -0.5000 &  -0.5000 &   0.99429283 &   0.99821574 &  239 &  2.00 &  -0.3713 &   0.8500 &   0.98618919 &   0.99841454\\
 240 &  2.00 &  -0.3000 &  -0.3000 &   0.99752736 &   0.99769231 &  241 &  2.00 &  -0.3000 &   0.0000 &   0.99884598 &   0.99869134\\
 242 &  2.00 &  -0.3000 &   0.3000 &   0.99610240 &   0.99918985 &  243 &  2.00 &  -0.1287 &  -0.8500 &   0.99304984 &   0.99007681\\
 244 &  2.00 &  -0.0000 &  -0.6000 &   0.99516204 &   0.98619699 &  245 &  2.00 &   0.0000 &  -0.3000 &   0.99892545 &   0.99242227\\
 246 &  2.00 &   0.0000 &   0.3000 &   0.99863224 &   0.99731761 &  247 &  2.00 &  -0.0000 &   0.6000 &   0.99634506 &   0.99856086\\
 248 &  2.00 &   0.1287 &   0.8500 &   0.99361958 &   0.99900743 &  249 &  2.00 &   0.3000 &  -0.3000 &   0.99662327 &   0.99533609\\
 250 &  2.00 &   0.3000 &   0.0000 &   0.99797332 &   0.99788987 &  251 &  2.00 &   0.3000 &   0.3000 &   0.99876214 &   0.99787514\\
 252 &  2.00 &   0.3713 &  -0.8500 &   0.98827818 &   0.98467140 &  253 &  2.00 &   0.5000 &   0.5000 &   0.98941751 &   0.99825143\\
 254 &  2.00 &   0.6000 &  -0.6000 &   0.96984460 &   0.99810394 &  255 &  2.00 &   0.6000 &  -0.0000 &   0.95948464 &   0.99845699\\
 256 &  2.00 &   0.6000 &   0.6000 &   0.92832500 &   0.99740142 &  257 &  2.00 &   0.8500 &   0.8500 &  FAIL &   0.99774880\\
 258 &  2.00 &   0.8713 &  -0.8500 &   0.89926461 &   0.99353572 &  259 &  2.50 &   0.0000 &   0.0000 &   0.99970931 &   0.99537931\\
 260 &  3.00 &  -0.8500 &  -0.8500 &  FAIL &   0.98436420 &  261 &  3.00 &  -0.7313 &   0.8500 &  FAIL &   0.99735308\\
 262 &  3.00 &  -0.6000 &   0.0000 &   0.99497798 &   0.99900780 &  262 &  3.00 &  -0.6000 &   0.0000 &   0.99497798 &   0.99900780\\
 263 &  3.00 &  -0.6000 &   0.6000 &   0.98886320 &   0.99887137 &  263 &  3.00 &  -0.6000 &   0.6000 &   0.98886320 &   0.99887137\\
 264 &  3.00 &  -0.6000 &  -0.6000 &  FAIL &   0.99721967 &  264 &  3.00 &  -0.6000 &  -0.6000 &  FAIL &   0.99721967\\
 265 &  3.00 &  -0.6000 &  -0.4000 &   0.99764488 &   0.99851513 &  265 &  3.00 &  -0.6000 &  -0.4000 &   0.99764488 &   0.99851513\\
 266 &  3.00 &  -0.6000 &   0.4000 &   0.99126457 &   0.99927475 &  267 &  3.00 &  -0.5000 &  -0.5000 &   0.99834314 &   0.99788203\\
 268 &  3.00 &  -0.4000 &  -0.6000 &   0.99829862 &   0.99535591 &  269 &  3.00 &  -0.4000 &   0.6000 &   0.98991667 &   0.99865836\\
 270 &  3.00 &  -0.3000 &  -0.3000 &   0.99912281 &   0.99770108 &  271 &  3.00 &  -0.3000 &   0.0000 &   0.99769789 &   0.99908172\\
 272 &  3.00 &  -0.3000 &   0.3000 &   0.99450676 &   0.99951495 &  273 &  3.00 &  -0.2687 &  -0.8500 &   0.99755264 &   0.99002444\\
 274 &  3.00 &  -0.2313 &   0.8500 &   0.98657464 &   0.99851515 &  275 &  3.00 &  -0.0000 &  -0.6000 &   0.99630503 &   0.98678490\\
 276 &  3.00 &   0.0000 &  -0.3000 &   0.99848126 &   0.99414093 &  277 &  3.00 &   0.0000 &   0.3000 &   0.99825834 &   0.99820955\\
 278 &  3.00 &  -0.0000 &   0.6000 &   0.99533306 &   0.99873319 &  279 &  3.00 &   0.2313 &  -0.8500 &   0.98840909 &   0.98392488\\
 280 &  3.00 &   0.2687 &   0.8500 &   0.99835881 &   0.99539385 &  281 &  3.00 &   0.3000 &  -0.3000 &   0.98920349 &   0.99681550\\
 282 &  3.00 &   0.3000 &  -0.0000 &   0.99176764 &   0.99833902 &  283 &  3.00 &   0.3000 &   0.3000 &   0.99374968 &   0.99684091\\
 284 &  3.00 &   0.4000 &  -0.6000 &   0.96967648 &   0.99538362 &  285 &  3.00 &   0.4000 &   0.6000 &   0.98240604 &   0.99578757\\
 286 &  3.00 &   0.5000 &   0.5000 &   0.94390756 &   0.99676160 &  287 &  3.00 &   0.6000 &  -0.6000 &   0.84455102 &   0.99835912\\
 288 &  3.00 &   0.6000 &  -0.4000 &   0.82742072 &   0.99828108 &  289 &  3.00 &   0.6000 &  -0.0000 &   0.76717729 &   0.99688503\\
 290 &  3.00 &   0.6000 &   0.4000 &   0.66855977 &   0.99800855 &  291 &  3.00 &   0.6000 &   0.6000 &   0.63311447 &   0.99797603\\
 292 &  3.00 &   0.7313 &  -0.8500 &   0.34131486 &   0.99882880 &  293 &  3.00 &   0.8500 &   0.8500 &  FAIL &   0.99413177\\
 294 &  3.50 &   0.0000 &   0.0000 &   0.99962744 &   0.99720619 &  295 &  4.50 &   0.0000 &   0.0000 &   0.99959992 &   0.99852620\\
 296 &  5.50 &   0.0000 &   0.0000 &   0.99948859 &   0.99917213 &  297 &  6.50 &   0.0000 &   0.0000 &   0.99954862 &   0.99918357\\
 298 &  7.00 &   0.0000 &   0.0000 &   0.99893013 &   0.99779383 &  299 &  7.50 &   0.0000 &  -0.0000 &   0.99917746 &   0.99905527\\
 300 &  8.50 &   0.0000 &  -0.0000 &   0.99809225 &   0.99795980 &  301 &  9.00 &   0.0000 &  -0.0000 &   0.99915872 &   0.99883787\\
 302 &  9.50 &   0.0000 &  -0.0000 &   0.99872859 &   0.99575710 &  302 &  9.50 &   0.0000 &  -0.0000 &   0.99872859 &   0.99575710\\
 303 & 10.00 &  -0.0000 &  -0.0000 &   0.99914437 &   0.99858701 &  303 & 10.00 &  -0.0000 &  -0.0000 &   0.99914437 &   0.99858701\\
 304 &  1.00 &   0.5000 &  -0.5000 &   0.99820189 &   0.99883669 &  304 &  1.00 &   0.5000 &  -0.5000 &   0.99820189 &   0.99883669\\
 305 &  1.22 &   0.3300 &  -0.4400 &   0.99935541 &   0.99562220 &  305 &  1.22 &   0.3300 &  -0.4400 &   0.99935541 &   0.99562220\\
 306 &  1.31 &   0.9617 &  -0.8997 &   0.99407715 &   0.99388921 &  306 &  1.31 &   0.9617 &  -0.8997 &   0.99407715 &   0.99388921\\
 307 &  1.23 &   0.3200 &  -0.5800 &   0.99557988 &   0.99518889 & 1419 &  8.00 &  -0.8000 &  -0.8000 &  FAIL &   0.99601120\\
1420 &  8.00 &  -0.8000 &   0.8000 &  FAIL &   0.99291147 & 1421 &  7.81 &  -0.6081 &   0.8000 &  FAIL &   0.98991770\\
1422 &  7.95 &  -0.8000 &  -0.4590 &  FAIL &   0.99304310 & 1423 &  8.00 &  -0.6035 &  -0.7532 &  FAIL &   0.99372499\\
1424 &  6.46 &  -0.6569 &  -0.8000 &  FAIL &   0.99087967 & 1425 &  6.12 &  -0.8000 &   0.6730 &  FAIL &   0.99326732\\
1426 &  8.00 &   0.4838 &   0.7484 &   0.81438685 &   0.98350230 & 1427 &  7.41 &  -0.6102 &  -0.7317 &  FAIL &   0.99181130\\
1428 &  5.52 &  -0.8000 &  -0.6998 &  FAIL &   0.99166892 & 1429 &  7.75 &  -0.2003 &  -0.7797 &   0.99728210 &   0.99164391\\
1430 &  8.00 &   0.2843 &  -0.7522 &   0.98071601 &   0.99497805 & 1431 &  8.00 &   0.0778 &  -0.7768 &   0.99838232 &   0.99283198\\
1432 &  5.84 &   0.6577 &   0.7930 &   0.48204071 &   0.98411561 & 1433 &  8.00 &  -0.7371 &   0.2075 &  FAIL &   0.99212739\\
1434 &  4.37 &   0.7977 &   0.7959 &  FAIL &   0.99224256 & 1435 &  6.59 &  -0.7895 &   0.0673 &  FAIL &   0.99059434\\
1436 &  6.28 &   0.0095 &  -0.8000 &   0.99869989 &   0.99008727 & 1437 &  6.04 &   0.8000 &   0.1476 &  FAIL &   0.99093498\\
1438 &  5.87 &   0.1257 &   0.8000 &   0.99734774 &   0.99748334 & 1439 &  6.48 &   0.7206 &  -0.3193 &   0.42259485 &   0.99479682\\
1440 &  5.64 &   0.7698 &   0.3065 &  FAIL &   0.99183894 & 1441 &  8.00 &   0.5955 &  -0.4781 &   0.58304414 &   0.99072892\\
1442 &  6.58 &  -0.7099 &  -0.1786 &  FAIL &   0.99586717 & 1443 &  5.68 &   0.4081 &  -0.7377 &   0.93185754 &   0.99214066\\
\hline
\end{tabular}
\end{table*}
\setcounter{table}{0}
\begin{table*}
\centering
\caption{\textit{(Continued)}}
\begin{tabular}{p{0.75cm}<\centering |p{0.8cm}<\centering |p{1.2cm}<\centering |p{1.2cm}<\centering |p{2.0cm}<\centering | p{2.0cm}<\centering || p{0.75cm}<\centering |p{0.8cm}<\centering |p{1.2cm}<\centering |p{1.2cm}<\centering | p{2.0cm}<\centering |p{2.0cm}<\centering}
\hline
ID & $q$ & $\chi_1$ & $\chi_2$ & v1FF & v4FF & ID & $q$ & $\chi_1$ & $\chi_2$ & v1FF & v4FF \\
\hline
1444 &  5.94 &  -0.0630 &  -0.7589 &   0.99459368 &   0.99091162 & 1445 &  4.67 &  -0.4971 &   0.8000 &   0.98810982 &   0.99456634\\
1446 &  3.15 &  -0.8000 &   0.7770 &  FAIL &   0.99854387 & 1447 &  3.16 &   0.7398 &   0.8000 &  FAIL &   0.99869873\\
1448 &  6.94 &  -0.4816 &   0.5248 &   0.99266244 &   0.99732545 & 1449 &  4.19 &  -0.8000 &  -0.3445 &  FAIL &   0.99338723\\
1450 &  4.07 &  -0.2836 &  -0.8000 &   0.99344190 &   0.98943940 & 1451 &  4.06 &   0.3133 &  -0.8000 &   0.97510875 &   0.99109158\\
1452 &  3.64 &   0.8000 &  -0.4266 &  FAIL &   0.99728541 & 1453 &  2.35 &   0.8000 &  -0.7845 &   0.87309598 &   0.99757038\\
1453 &  2.35 &   0.8000 &  -0.7845 &   0.87309598 &   0.99757038 & 1454 &  2.45 &  -0.8000 &  -0.7340 &  FAIL &   0.99283404\\
1454 &  2.45 &  -0.8000 &  -0.7340 &  FAIL &   0.99283404 & 1455 &  8.00 &  -0.3978 &   0.0014 &   0.99535512 &   0.99818015\\
1455 &  8.00 &  -0.3978 &   0.0014 &   0.99535512 &   0.99818015 & 1456 &  3.00 &   0.7449 &   0.6970 &  FAIL &   0.99821791\\
1456 &  3.00 &   0.7449 &   0.6970 &  FAIL &   0.99821791 & 1457 &  3.25 &   0.5448 &   0.8000 &   0.85339194 &   0.99700079\\
1457 &  3.25 &   0.5448 &   0.8000 &   0.85339194 &   0.99700079 & 1458 &  3.80 &  -0.0634 &   0.8000 &   0.99093055 &   0.99507874\\
1458 &  3.80 &  -0.0634 &   0.8000 &   0.99093055 &   0.99507874 & 1459 &  2.26 &   0.7632 &   0.8000 &  FAIL &   0.99407070\\
1460 &  8.00 &   0.1236 &   0.1087 &   0.99812093 &   0.99838676 & 1461 &  2.88 &  -0.4486 &  -0.8000 &   0.99407518 &   0.99108842\\
1462 &  2.63 &  -0.8000 &   0.5107 &  FAIL &   0.99896681 & 1463 &  4.98 &   0.6129 &   0.2407 &   0.59466503 &   0.98902720\\
1464 &  6.54 &  -0.0522 &  -0.3226 &   0.99868326 &   0.99824380 & 1465 &  1.71 &  -0.7875 &   0.7674 &   0.98971051 &   0.99210525\\
1466 &  1.90 &   0.6988 &  -0.8000 &   0.94899160 &   0.99784275 & 1467 &  2.23 &  -0.5622 &   0.8000 &   0.98033831 &   0.99776091\\
1468 &  2.27 &   0.5147 &   0.8000 &   0.96789595 &   0.99365138 & 1469 &  1.85 &   0.8000 &   0.6689 &  FAIL &   0.99709673\\
1470 &  1.52 &  -0.7286 &  -0.7896 &  FAIL &   0.99896011 & 1471 &  1.33 &  -0.7754 &  -0.8000 &  FAIL &   0.99907051\\
1472 &  2.37 &  -0.8000 &  -0.1160 &  FAIL &   0.99925827 & 1473 &  1.45 &   0.6976 &   0.7879 &  FAIL &   0.99706091\\
1474 &  1.28 &   0.7239 &  -0.8000 &   0.99804985 &   0.99634305 & 1475 &  1.00 &  -0.8000 &  -0.8000 &  FAIL &   0.99907391\\
1476 &  1.00 &  -0.8000 &   0.8000 &   0.99749323 &   0.99851389 & 1477 &  1.00 &   0.8000 &   0.8000 &  FAIL &   0.99663564\\
1478 &  1.97 &   0.8000 &   0.1263 &   0.58061589 &   0.99424776 & 1479 &  1.55 &  -0.5569 &  -0.8000 &   0.99742965 &   0.99955394\\
1480 &  1.55 &  -0.8000 &  -0.3097 &   0.99754270 &   0.99690872 & 1481 &  1.00 &   0.7312 &   0.7938 &  FAIL &   0.99675088\\
1482 &  1.39 &  -0.5814 &   0.8000 &   0.99241046 &   0.99872451 & 1483 &  3.17 &   0.5605 &  -0.1940 &   0.89561085 &   0.99213719\\
1484 &  2.90 &  -0.5603 &   0.2973 &   0.99028412 &   0.99865419 & 1485 &  3.09 &   0.3526 &  -0.4006 &   0.97868331 &   0.99602105\\
1486 &  3.72 &   0.4273 &  -0.0335 &   0.95146127 &   0.99554587 & 1487 &  1.25 &  -0.8000 &   0.5117 &   0.99657160 &   0.99742029\\
1488 &  1.59 &  -0.3277 &   0.7475 &   0.99192608 &   0.99901794 & 1489 &  3.46 &   0.2999 &  -0.1717 &   0.98650941 &   0.99817189\\
1490 &  1.25 &   0.4121 &   0.7595 &   0.98698538 &   0.99732650 & 1491 &  1.66 &   0.1991 &  -0.6966 &   0.99633681 &   0.99053897\\
1492 &  1.00 &  -0.4707 &  -0.7936 &   0.99783519 &   0.99910379 & 1493 &  1.28 &   0.0098 &   0.8000 &   0.99179755 &   0.99820165\\
1494 &  2.21 &  -0.4743 &  -0.3891 &   0.99685568 &   0.99703984 & 1495 &  1.00 &   0.7792 &   0.5336 &   0.93354574 &   0.99624815\\
1496 &  1.16 &   0.8000 &   0.0286 &   0.99724171 &   0.99128659 & 1497 &  1.00 &   0.6824 &   0.6676 &   0.90943769 &   0.99693165\\
1498 &  1.03 &   0.2233 &  -0.7838 &   0.99780380 &   0.99759788 & 1499 &  1.00 &  -0.7547 &   0.3429 &   0.99831812 &   0.99752713\\
\hline
\end{tabular}
\end{table*}

\begin{figure*}
\begin{tabular}{cc}
\includegraphics[width=0.5\textwidth]{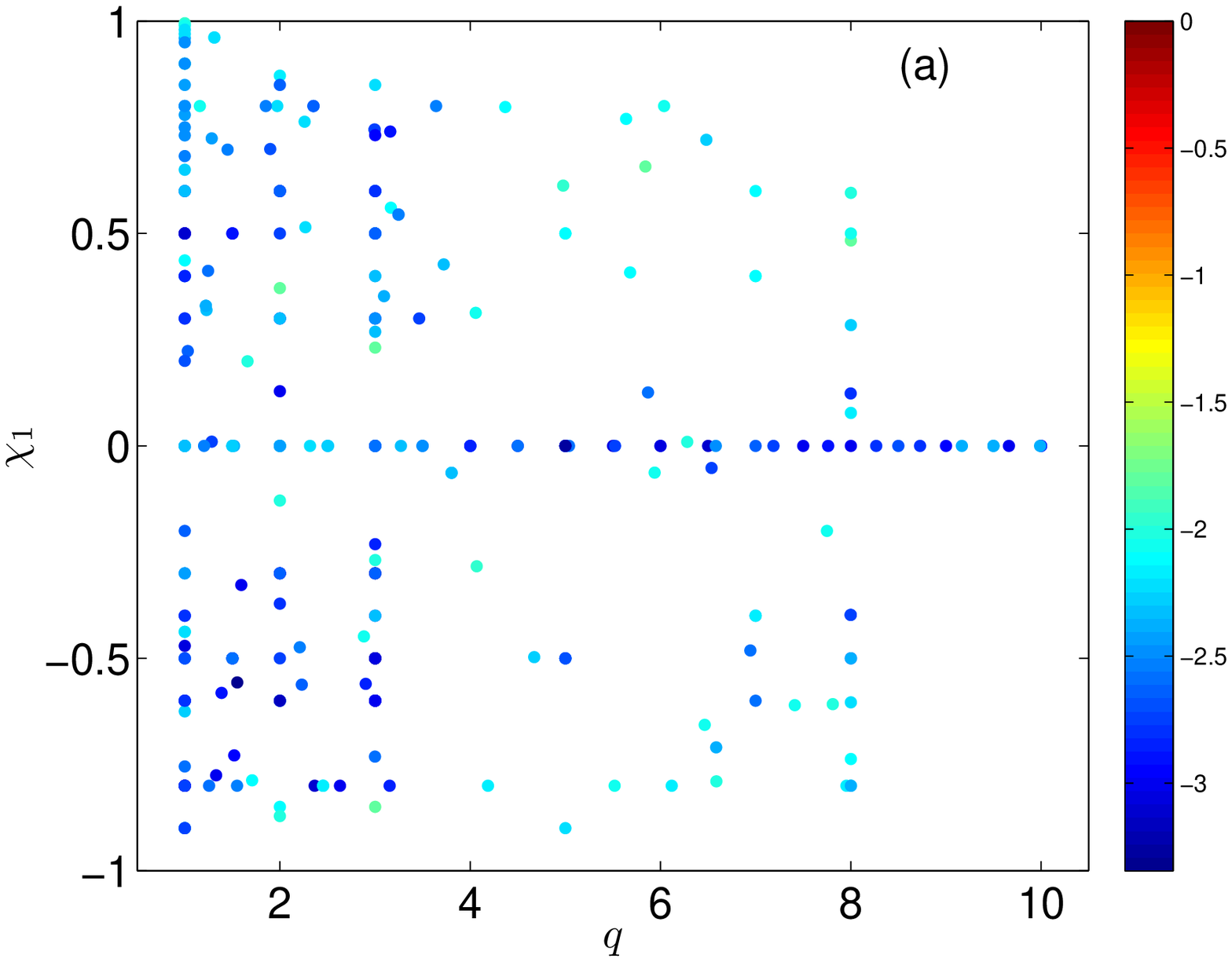}&
\includegraphics[width=0.5\textwidth]{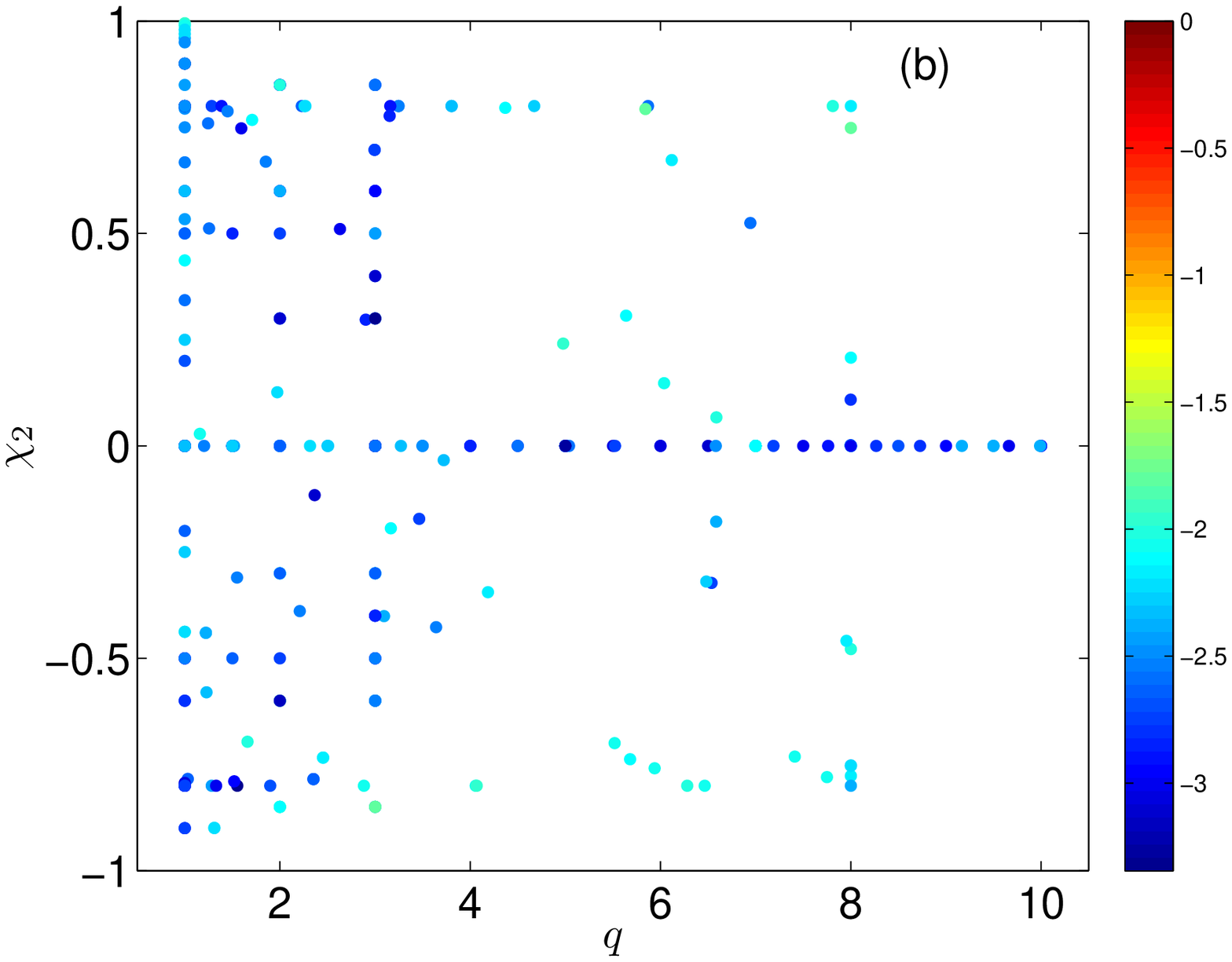}
\end{tabular}
\caption{The SXS simulation waveforms with small eccentricity used to validate SEOBNRE models. We project the 3D parameter space of spinning, nonprecessing waveforms to the mass ratio $q=m_1/m_2>1$ and the two dimensionless BH spin $\chi_{1,2}$. The color represents $\log_{10}(1-{\rm FF})$.}\label{fig2}
\end{figure*}
\begin{figure*}
\begin{tabular}{cc}
\includegraphics[width=0.49\textwidth]{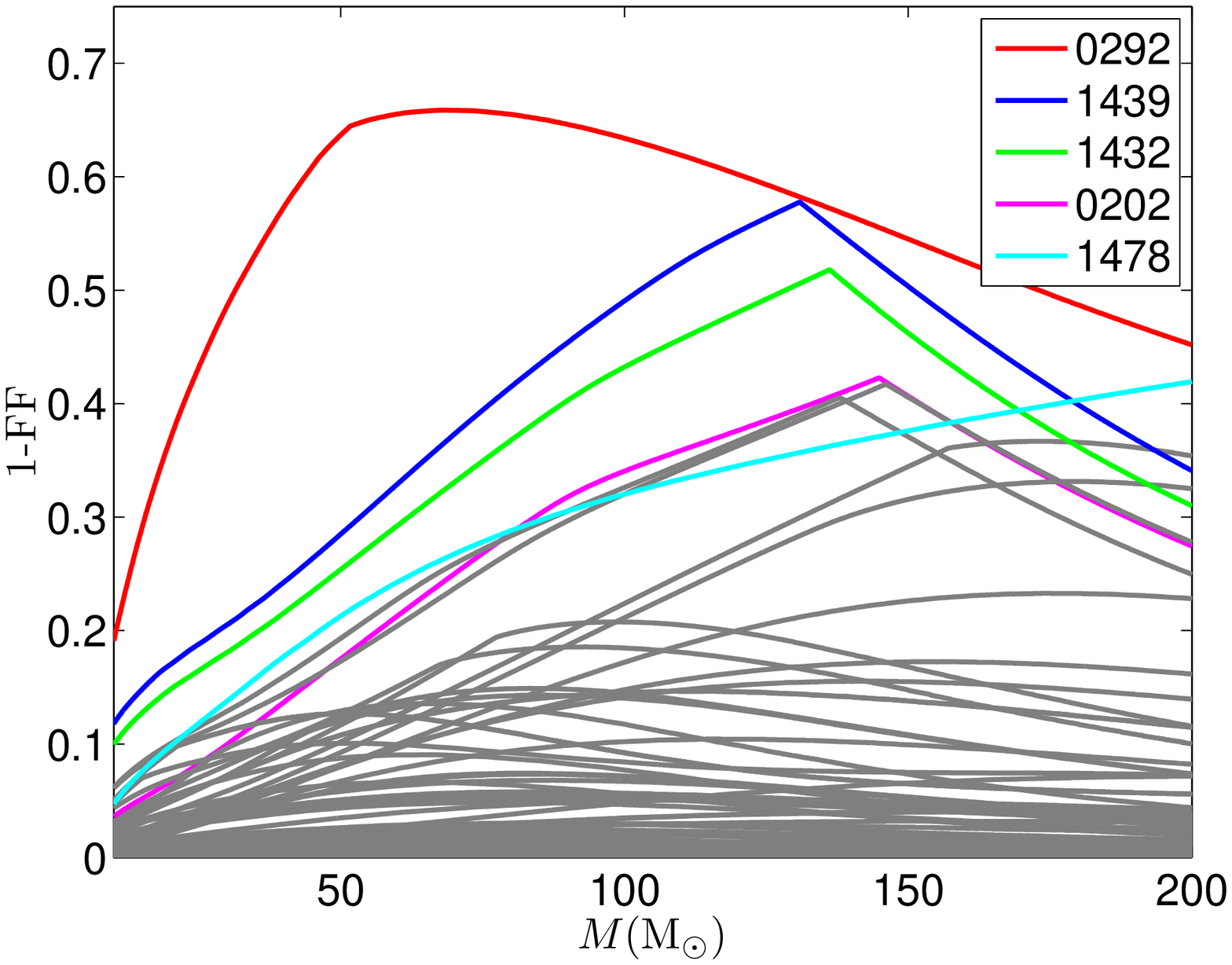}&
\includegraphics[width=0.51\textwidth]{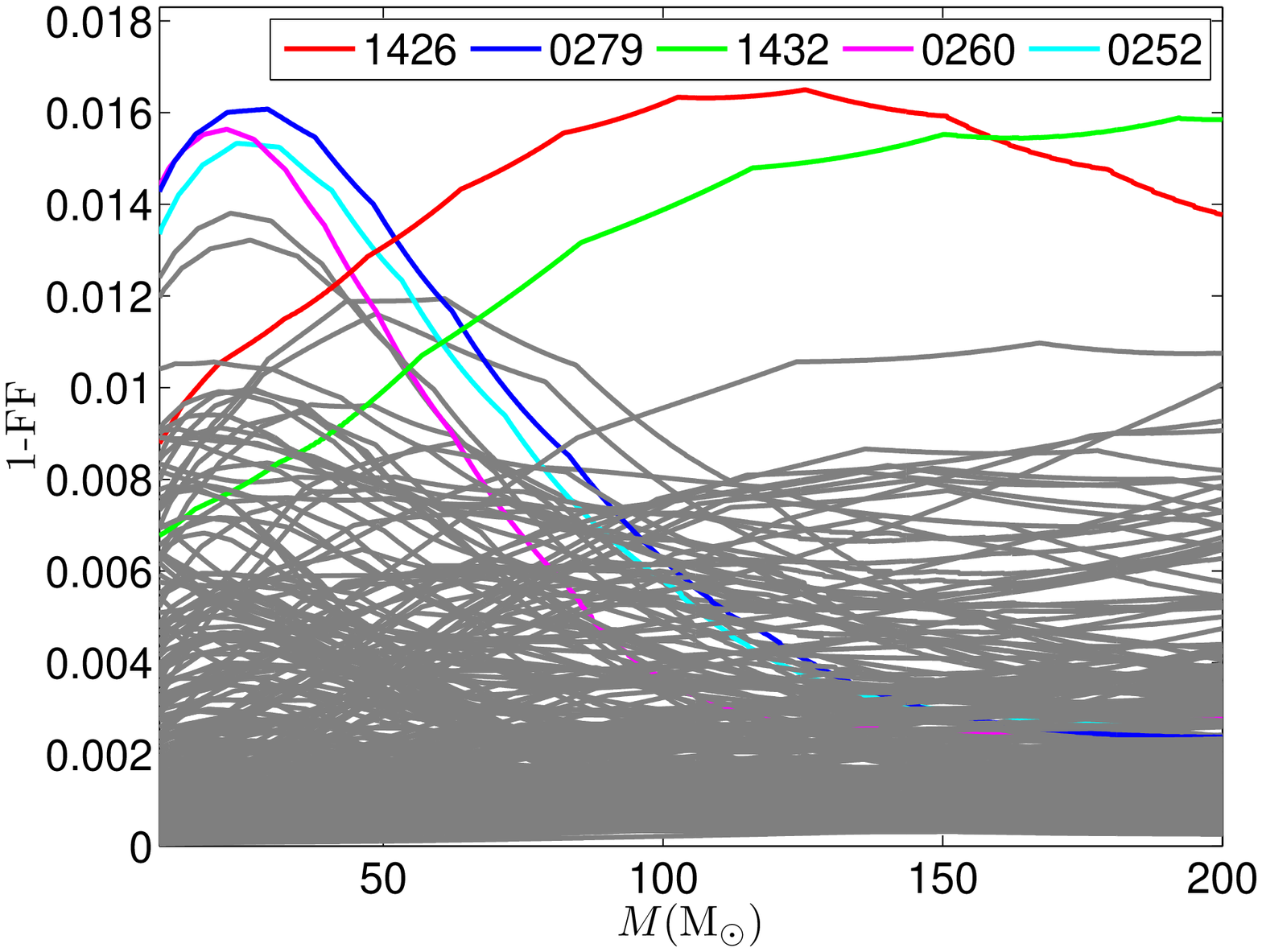}
\end{tabular}
\caption{Validating SEOBNRE waveform models to SXS simulations with small eccentricity. The left panel is for SEOBNREv1 and the right panel is for SEOBNREv4. The legend shows the ID of SXS simulations. There are 300 lines in each panel. Only 5 lines with smallest fitting factor are marked out with legend in each panel.}\label{fig3}
\end{figure*}
Respect to the advanced LIGO and the advanced Virgo, we define inner product of two given waveforms $h_1(t)$ and $h_2(t)$ as
\begin{align}
(h_1|h_2)\equiv 4\Re\int_{f_{\rm min}}^{f_{\rm max}}\frac{\tilde{h}_1(f)\tilde{h}^*_2(f)}{S_n(f)}df,\label{equation2}
\end{align}
where $S_n(f)$ is the one-sided power spectral density of the detector's noise. $(f_{\rm min},f_{\rm max})$ corresponds to the detector frequency band. In the current paper we use the designed sensitivity of advanced LIGO \cite{Sho10}. Respectively $f_{\rm min} = 10$Hz and $f_{\rm max} = 8192$Hz.
\begin{figure}
\begin{tabular}{c}
\includegraphics[width=0.5\textwidth]{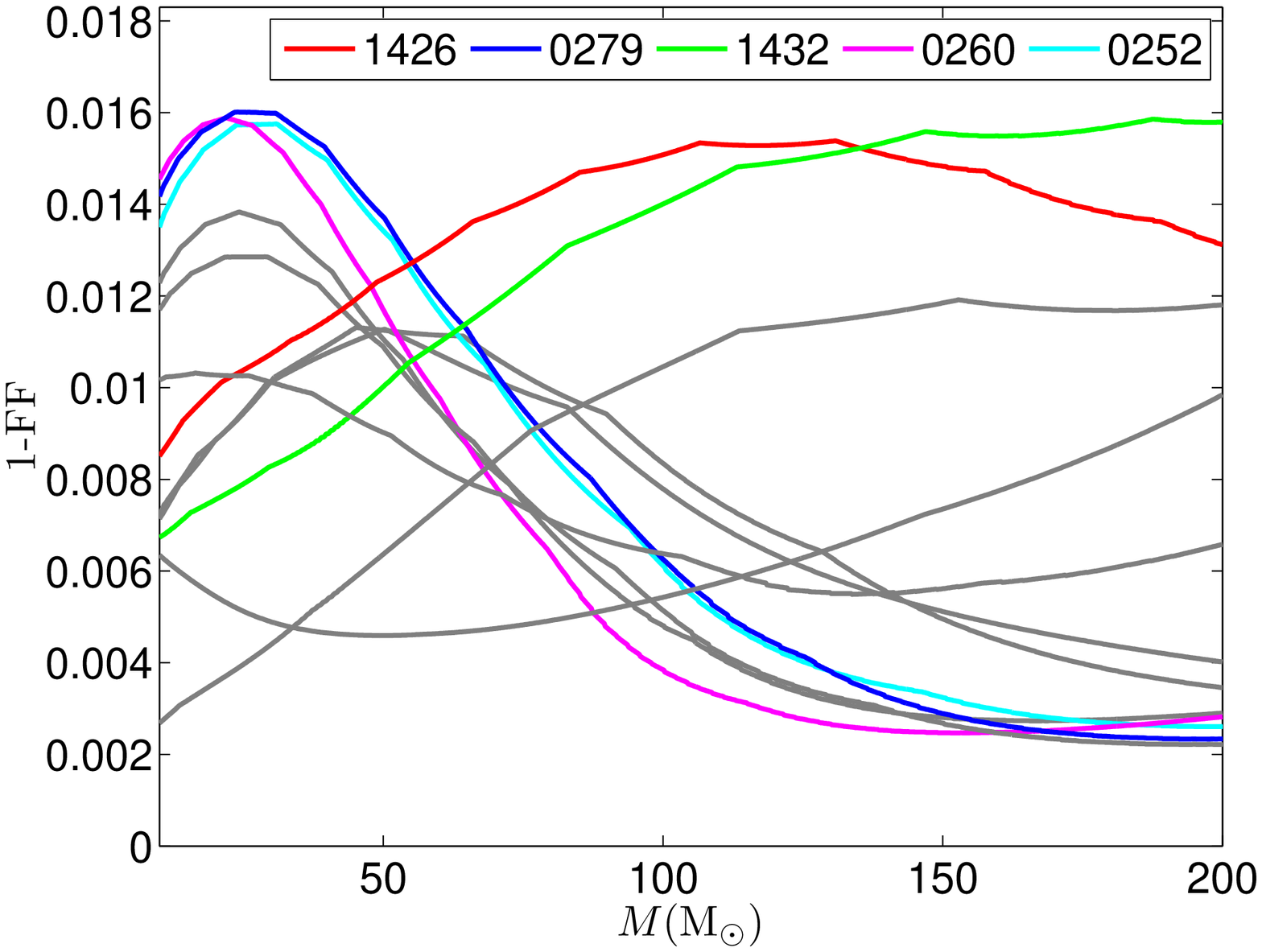}
\end{tabular}
\caption{Comparing the SXS simulation waveforms to SEOBNRv4 waveforms for the 12 cases with fitting factor smaller than 99\% shown in the right panel of the Fig.~\ref{fig3}. The plot convention is the same to that of the Fig.~\ref{fig3}.}\label{fig4}
\end{figure}

\begin{figure*}
\begin{tabular}{cc}
\includegraphics[width=0.495\textwidth]{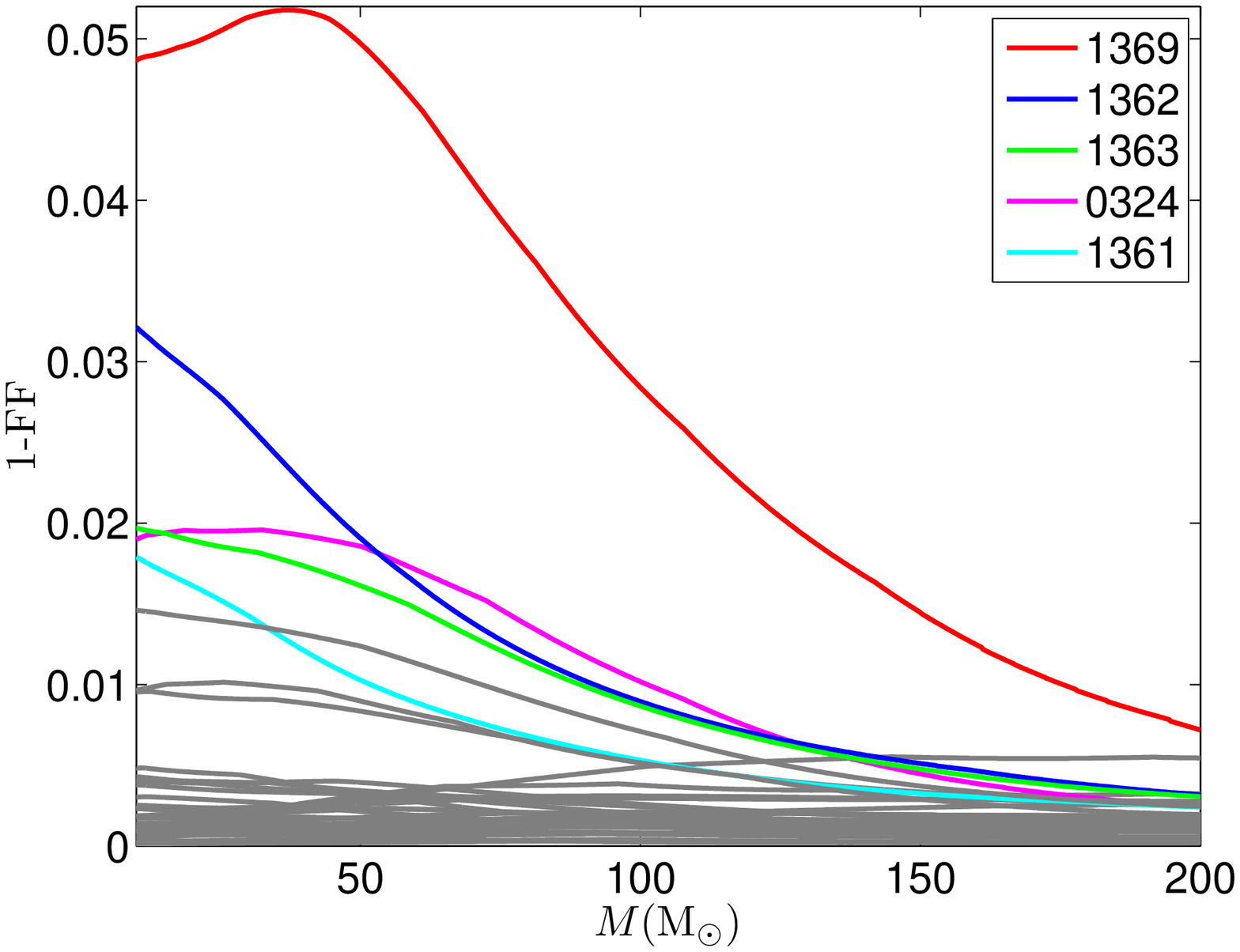}&
\includegraphics[width=0.505\textwidth]{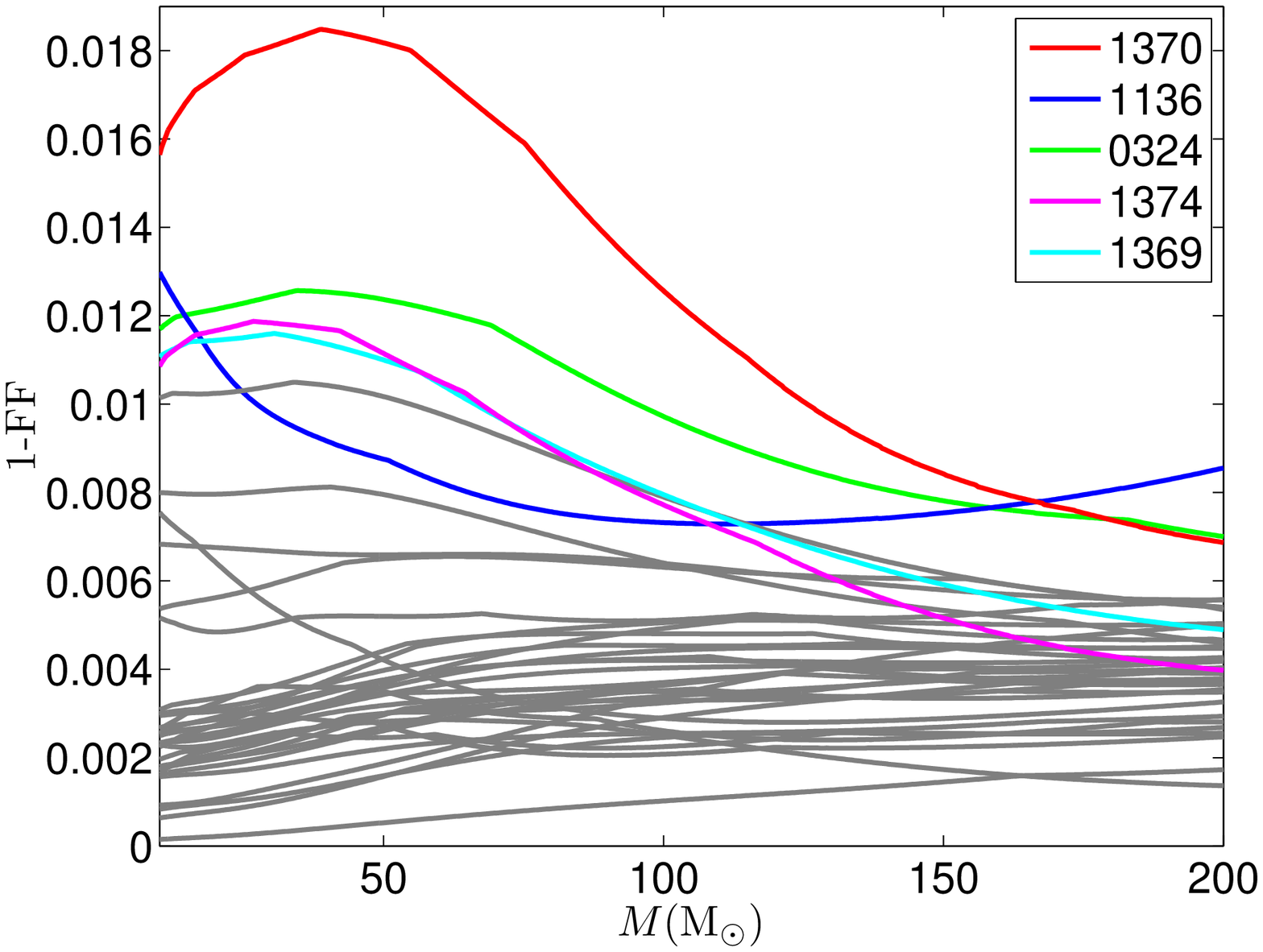}
\end{tabular}
\caption{Validating SEOBNRE waveform models to eccentric SXS simulations. The left panel is for SEOBNREv1 and the right panel is for SEOBNREv4. The legend shows the ID of SXS simulations. There are 32 lines in the left panel. And there are 35 lines in the right panel. The 3 missing cases in the left panel correspond to the failed cases for SEOBNREv1. Only 5 lines with smallest fitting factor are marked out with legend in each panel.}\label{fig5}
\end{figure*}
\begin{figure*}
\begin{tabular}{c}
\includegraphics[width=\textwidth]{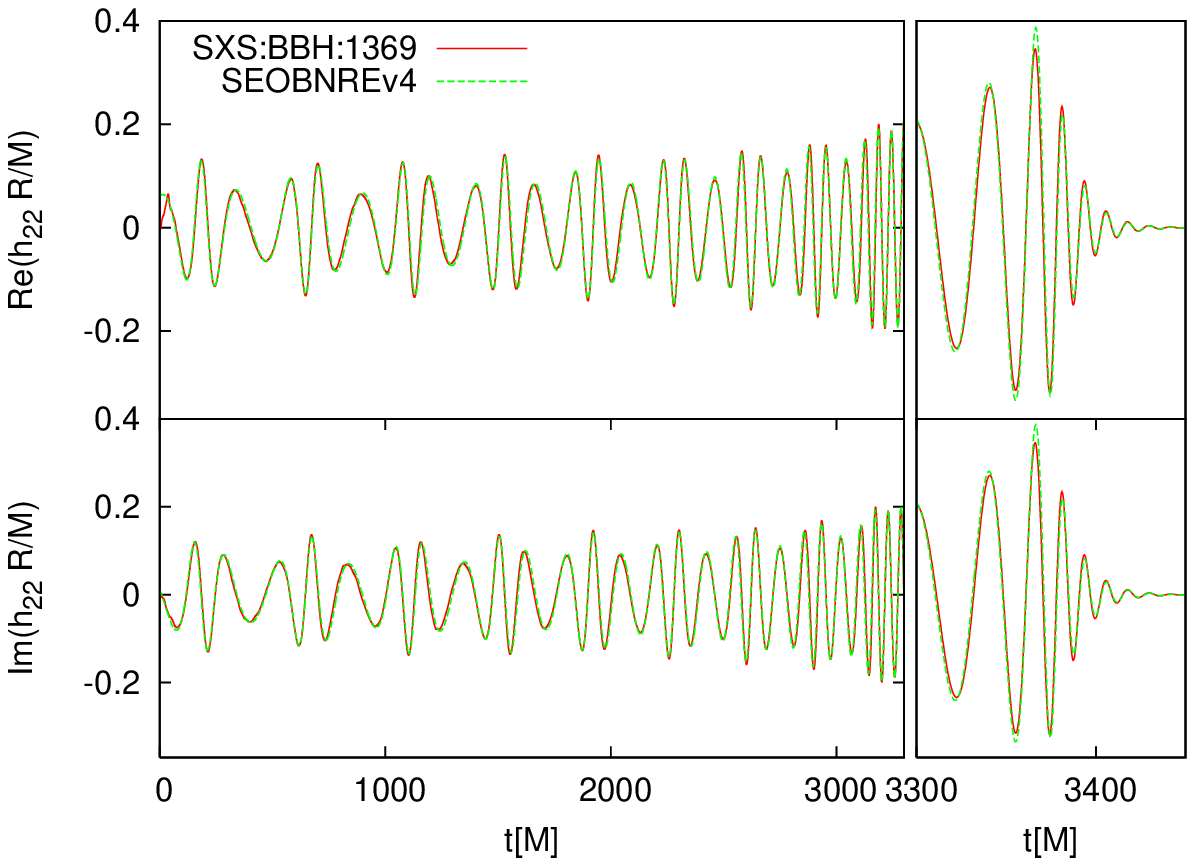}
\end{tabular}
\caption{Waveform comparison between NR and the SEOBNREv4 model for SXS:BBH:1369. The corresponding fitting factor is ${\rm FF}=98.8\%$. The initial eccentricity at the reference frequency $Mf_0\approx0.002$ is $e_0=0.59$ estimated by the SEOBNREv4 model.}\label{fig6}
\end{figure*}
\subsection{Small eccentricity cases}
There are 300 small eccentricity cases among the SXS catalogue \cite{SXSBBH}. The eccentricities of these simulations are less than 0.01 when the numerical simulation starts.
The parameters including mass ratio $q$ and spin $\chi$ of these 300 cases are shown in the Fig.~\ref{fig2}. We setup the reference frequency $f_0$ to the frequency when the numerical simulation starts and initial eccentricity $e_0$ according to the value provided by numerical relativity estimation.

The SEOBNREv1 model breaks down for some cases with large spin. There are 56 such failed cases. Among the rest 244 cases, the case with the smallest fitting factor is 0292 with FF 31.8\%. There are 65 cases among the 244 cases for SEOBNREv1 admitting ${\rm FF}<99\%$. There are 23 cases among these 65 cases admitting ${\rm FF}<90\%$.

In contrast, SEOBNREv4 model never breaks down. It works quite well for all 300 cases. There are only 12 cases for SEOBNREv4 with ${\rm FF} < 99\%$ among 300 cases in all. Even the smallest fitting factor among these 300 cases can reach 98.3\% corresponding to the case 1426, which admits eccentricity 0.0003326 when the numerical simulation starts.

Comparing the right panel to the left panel of the Fig.~\ref{fig3}, we can see that SEOBNREv4 improves quite much than SEOBNREv1. The fitting factors in the right panel of the Fig.~2 of \cite{PhysRevD.95.044028} are always bigger than 99\% is because the above mentioned 12 cases are all absent in the right panel of the Fig.~2 of \cite{PhysRevD.95.044028}. These 12 cases were not available when SEOBNRv4 was constructed \cite{PhysRevD.95.044028}. For comparison usage, we calculate these 12 cases for SEOBNRv4 waveform model in the Fig.~\ref{fig4}. We can see the behavior of SEOBNREv4 is roughly the same to that of SEOBNRv4 for taking them as quasi-circular cases.

The comparison results between SEOBNREv1 and SEOBNREv4 for the 300 cases are listed in the Table.~\ref{table1}. From these listed results we can see SEOBNREv1 fits better than SEOBNREv4 for some slow spinning binary black holes ($\chi_{1,2}<0.5$). But the difference is always less than 1\%. In another words, up to 1\%, the SEOBNREv4 behaves better than SEOBNREv1 for these small eccentricity cases.
\subsection{Large eccentricity cases}
\begin{table*}
\centering
\caption{Validating the SEOBNRE models against the SXS simulations with significant eccentricity. Notation convention is the same to that of Tab.~\ref{table1}. In addition, $e_{\rm v1}$ and $e_{\rm v4}$ are the determined eccentricity at reference frequency $Mf_0=0.002$ by SEOBNREv1 and SEOBNREv4 respectively. `--' for $e_{\rm v1}$ means not available.}
\begin{tabular}{p{0.7cm}<\centering |p{0.6cm}<\centering |p{1.1cm}<\centering |p{1.1cm}<\centering |p{1.3cm}<\centering|p{0.8cm}<\centering | p{1.3cm}<\centering|p{0.8cm}<\centering || p{0.7cm}<\centering |p{0.6cm}<\centering |p{1.1cm}<\centering |p{1.1cm}<\centering | p{1.3cm}<\centering|p{0.8cm}<\centering |p{1.3cm}<\centering|p{0.8cm}<\centering}
\hline
ID & $q$ & $\chi_1$ & $\chi_2$ & v1FF & $e_{\rm v1}$ & v4FF & $e_{\rm v4}$ & ID & $q$ & $\chi_1$ & $\chi_2$ & v1FF & $e_{\rm v1}$ & v4FF & $e_{\rm v4}$ \\
\hline
  83 & 1.00 &   0.5000 &   0.0000 &   0.99797 & 0.07 &   0.99500 & 0.07 &   87 & 1.00 &   0.0000 &   0.0000 &   0.99957 & 0.07 &   0.99732 & 0.03 \\
  89 & 1.00 &  -0.5000 &   0.0000 &   0.99704 & 0.15 &   0.99713 & 0.14 &   91 & 1.00 &   0.0000 &   0.0000 &   0.99966 & 0.05 &   0.99701 & 0.00 \\
 100 & 1.50 &   0.0000 &   0.0000 &   0.99960 & 0.06 &   0.99639 & 0.01 &  105 & 3.00 &  -0.5000 &   0.0000 &   0.99615 & 0.09 &   0.99827 & 0.06 \\
 106 & 5.00 &   0.0000 &  -0.0000 &   0.99739 & 0.12 &   0.99744 & 0.13 &  108 & 5.00 &  -0.5000 &  -0.0000 &   0.99447 & 0.13 &   0.99597 & 0.13 \\
 321 & 1.22 &   0.3300 &  -0.4400 &   0.99913 & 0.28 &   0.99597 & 0.24 &  322 & 1.22 &   0.3300 &  -0.4400 &   0.99874 & 0.32 &   0.99582 & 0.30 \\
 323 & 1.22 &   0.3300 &  -0.4400 &   0.99748 & 0.43 &   0.99479 & 0.42 &  324 & 1.22 &   0.3300 &  -0.4400 &   0.98042 & 0.58 &   0.98743 & 0.58 \\
1136 & 1.00 &  -0.7500 &  -0.7500 &  FAIL & -- &   0.98702 & 0.39 & 1149 & 3.00 &   0.7000 &   0.6000 &  FAIL & -- &   0.99245 & 0.02 \\
1169 & 3.00 &  -0.7000 &  -0.6000 &  FAIL & -- &   0.99442 & 0.12 & 1355 & 1.00 &  -0.0000 &  -0.0000 &   0.99690 & 0.29 &   0.99489 & 0.29 \\
1356 & 1.00 &  -0.0000 &  -0.0000 &   0.99517 & 0.34 &   0.99645 & 0.34 & 1357 & 1.00 &  -0.0000 &  -0.0000 &   0.99573 & 0.45 &   0.99571 & 0.43 \\
1358 & 1.00 &   0.0000 &  -0.0000 &   0.99598 & 0.44 &   0.99549 & 0.44 & 1359 & 1.00 &  -0.0000 &  -0.0000 &   0.99621 & 0.45 &   0.99589 & 0.45 \\
1360 & 1.00 &  -0.0000 &   0.0000 &   0.99043 & 0.53 &   0.99345 & 0.53 & 1361 & 1.00 &  -0.0000 &  -0.0000 &   0.98211 & 0.54 &   0.99317 & 0.53 \\
1362 & 1.00 &   0.0000 &  -0.0000 &   0.96786 & 0.60 &   0.99188 & 0.59 & 1363 & 1.00 &  -0.0000 &   0.0000 &   0.98032 & 0.59 &   0.98951 & 0.59 \\
1364 & 2.00 &  -0.0000 &  -0.0000 &   0.99925 & 0.29 &   0.99629 & 0.27 & 1365 & 2.00 &  -0.0000 &  -0.0000 &   0.99906 & 0.33 &   0.99598 & 0.33 \\
1366 & 2.00 &  -0.0000 &   0.0001 &   0.99767 & 0.45 &   0.99540 & 0.43 & 1367 & 2.00 &  -0.0000 &  -0.0000 &   0.99593 & 0.43 &   0.99573 & 0.43 \\
1368 & 2.00 &  -0.0000 &  -0.0000 &   0.99789 & 0.46 &   0.99519 & 0.44 & 1369 & 2.00 &   0.0000 &   0.0000 &   0.94822 & 0.59 &   0.98840 & 0.58 \\
1370 & 2.00 &   0.0000 &   0.0000 &   0.98540 & 0.60 &   0.98152 & 0.58 & 1371 & 3.00 &   0.0000 &  -0.0000 &   0.99695 & 0.28 &   0.99706 & 0.29 \\
1372 & 3.00 &  -0.0000 &  -0.0000 &   0.99819 & 0.45 &   0.99625 & 0.42 & 1373 & 3.00 &   0.0000 &  -0.0000 &   0.99883 & 0.46 &   0.99649 & 0.43 \\
1374 & 3.00 &   0.0000 &  -0.0000 &   0.98986 & 0.60 &   0.98813 & 0.58 &       &    &          &          &           &      &           &      \\
\hline
\end{tabular}\label{table2}
\end{table*}
Now we move on to the cases with significant eccentricities. The eccentricities of these simulations when the numerical simulation starts are bigger than 0.02 or even too big to be estimated through numerical relativity method. There are 35 significantly eccentric cases in the SXS catalog \cite{SXSBBH}. As we and others realized before \cite{PhysRevD.96.044028,PhysRevD.101.044049,PhysRevD.101.101501}, the initial eccentricity estimated by numerical relativity is not very meaningful. Consequently, the same procedure is done for the waveform comparison as in \cite{PhysRevD.96.044028,PhysRevD.101.044049} to search the corresponding parameter $e_0$. The determined eccentricities at reference frequency $Mf_0=0.002$ by SEOBNREv1 and SEOBNREv4 are consistent to each other. Note that $M$ is the total mass of the binary black hole, and $f_0\approx 40\frac{10{\rm M}_\odot}{M}$Hz. The difference of the estimated eccentricity between SEOBNREv1 and SEOBNREv4 is smaller than 0.05. The determined eccentricities fall in the range $e_0\lesssim0.6$. The corresponding procedure in \cite{PhysRevD.101.101501} is a little different, where both $f_0$ and $e_0$ are adjusted to align the waveforms between numerical relativity and the extended TEOBiResumS\_SM waveform.

The validating results are listed in the Table.~\ref{table2} and plotted in the Fig.~\ref{fig5}. For SEOBNREv1 there are 3 failed cases. And there are 7 cases admitting ${\rm FF}<99\%$. Among them there are two cases admitting ${\rm FF}<98\%$ and the worst case admits ${\rm FF}<95\%$. SEOBNREv4 works well for all 35 cases. And there are 6 cases admitting ${\rm FF}<99\%$. But there are only 3 cases admitting ${\rm FF}<98.8\%$. SXS:BBH:1370 is the most worst case which still admits fitting factor 98.8\%. When the total mass is bigger than 100M${}_\odot$, the fitting factor of SXS:BBH:1370 can also be larger than 99\%.

Corresponding to the time-domain waveform example shown in \cite{PhysRevD.101.101501}, we plot the waveform comparison between SEOBNREv4 and numerical relativity result in the Fig.~\ref{fig6} for significantly eccentric waveform without spin SXS:BBH:1369. Our fitting factor is 98.8\% and \cite{PhysRevD.101.101501} admits fitting factor 98.6\%. Again we need to note here we adjust only $e_0$ which is different to the procedure adjusting both $e_0$ and $f_0$ in \cite{PhysRevD.101.101501}. Regarding to the spinning eccentric waveforms, we have (\cite{PhysRevD.101.101501} has) 99.7\% (99.0\%) for SXS:BBH:89, 99.6\% (98.5\%) for SXS:BBH:321, 99.6\% (98.8\%) for SXS:BBH:322, 99.5\% (98.4\%) for SXS:BBH:323, 98.7\% (97.8\%) for SXS:BBH:324, 98.9\% (99.4\%) for SXS:BBH:1136, 99.2\% (96.8\%) for SXS:BBH:1149, 99.4\% (99.8\%) for SXS:BBH:1169.

Combining the results of Fig.~\ref{fig3} and \ref{fig5} we can see our SEOBNREv4 can always recover numerical relativity waveform with ${\rm FF}>98.2\%$ for parameters range $1\leq q\lesssim10$, $-0.9\lesssim\chi_{1,2}\lesssim0.995$ and $0\leq e_0\lesssim0.6$ at reference frequency $Mf_0=0.002$ (equivalently $f_0\approx 40\frac{10{\rm M}_\odot}{M}$Hz).

\section{SEOBNREv4 as a waveform model for supermassive binary black holes}
In this section we do a primary estimation of SEOBNREv4 as a waveform model for supermassive binary black holes. We consider LISA \cite{audley2017laser,PhysRevLett.118.171101}, Taiji \cite{Ruan2020} and Tianqin \cite{luo2016tianqin,Luo_2020} as example space-based detectors. We do not involve realistic response functions as \cite{2020arXiv200708544T}, instead we use sky averaged sensitivity \cite{Robson_2019} to do the estimation. And more the confusion noise is also ignored in the current estimation.

Specifically we use the following approximated sensitivity for space based gravitational wave detectors (Eq.~(13) of \cite{Robson_2019})
\begin{align}
S_n(f)&=\frac{10}{3L^2}\left(P_{\rm OMS}+2(1+\cos^2(f/f_*))\frac{P_{\rm acc}}{(2\pi f)^4}\right)\times\nonumber\\
&\left(1+\frac{6}{10}\left(\frac{f}{f_*}\right)^2\right),\\
f_*&=c/(2\pi L).
\end{align}
For LISA \cite{Robson_2019} we have
\begin{align}
P_{\rm OMS}&=(1.5\times10^{-11}{\rm m})^2{\rm Hz}^{-1},\\
P_{\rm acc}&=(3\times10^{-15}{\rm ms}^{-2})^2\left(1+\left(\frac{4\times10^{-4}{\rm Hz}}{f}\right)^2\right){\rm Hz}^{-1},\\
L&=2.5\times10^9{\rm m}.
\end{align}
For Taiji \cite{2018arXiv180709495R} we have
\begin{align}
P_{\rm OMS}&=(8\times10^{-11}{\rm m})^2{\rm Hz}^{-1},\\
P_{\rm acc}&=(3\times10^{-15}{\rm ms}^{-2})^2\left(1+\left(\frac{4\times10^{-4}{\rm Hz}}{f}\right)^2\right){\rm Hz}^{-1},\\
L&=3\times10^9{\rm m}.
\end{align}
For Tianqin we have \cite{luo2016tianqin}
\begin{align}
P_{\rm OMS}&=(1\times10^{-12}{\rm m})^2{\rm Hz}^{-1},\\
P_{\rm acc}&=(1\times10^{-15}{\rm ms}^{-2})^2\left(1+\left(\frac{1\times10^{-4}{\rm Hz}}{f}\right)^2\right){\rm Hz}^{-1},\\
L&=\sqrt{3}\times10^8{\rm m}.
\end{align}
Since the Taiji and Tianqin projects develop very fast, our approximated sensitivity curve maybe different to the ones shown in other publications (for example \cite{PhysRevD.100.044042,PhysRevD.102.024089}). But the above approximation is enough for the estimation usage needed in the current work.

\begin{figure}
\begin{tabular}{c}
\includegraphics[width=0.5\textwidth]{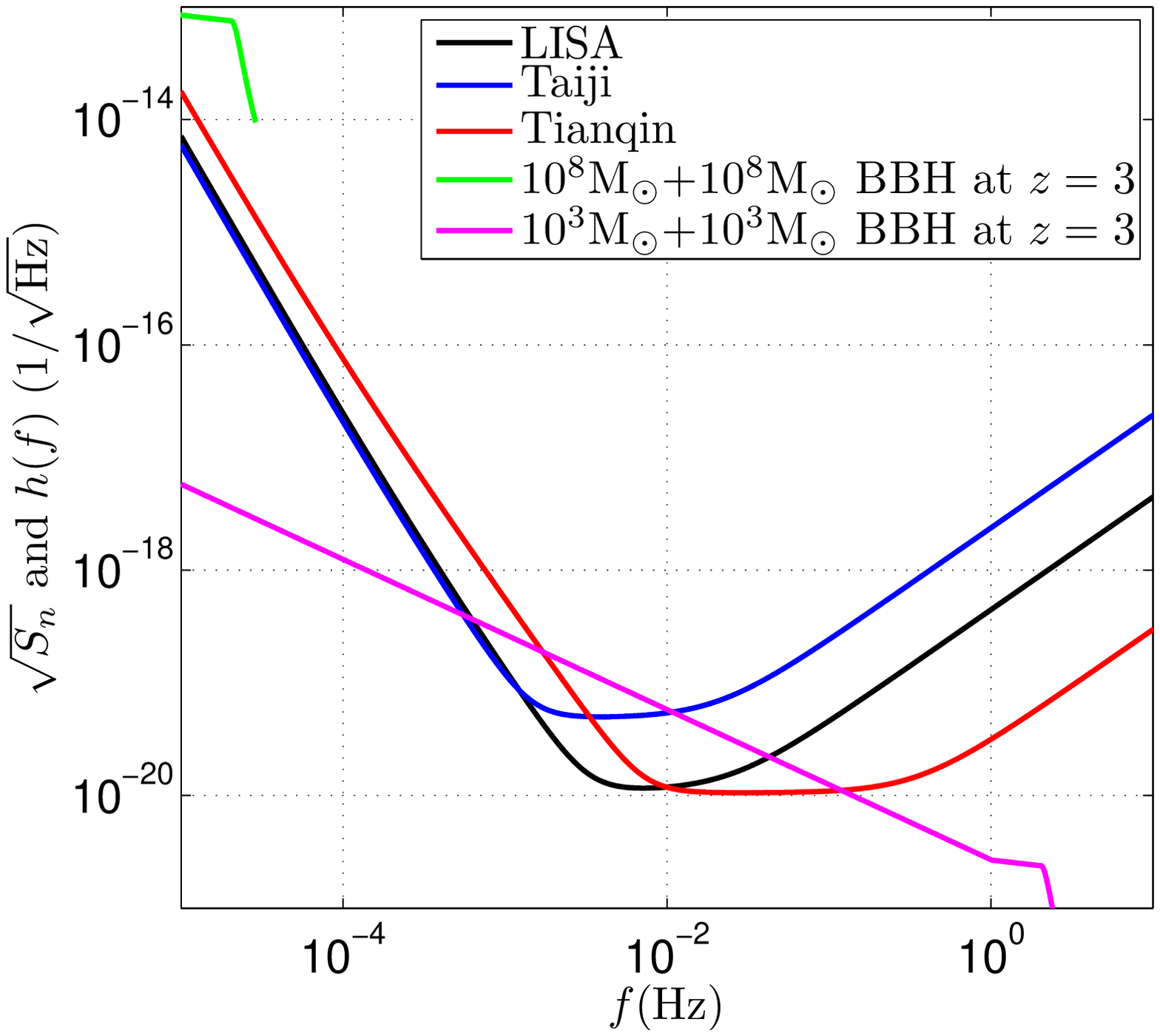}
\end{tabular}
\caption{Approximated sensitivity curves for LISA, Taiji and Tianqin. And approximated waveforms for two equal mass BBHs with source frame total mass $2\times10^8$M${}_\odot$ and $2\times10^3$M${}_\odot$ locating at $z=3$.}\label{fig7}
\end{figure}

\begin{figure}
\begin{tabular}{c}
\includegraphics[width=0.5\textwidth]{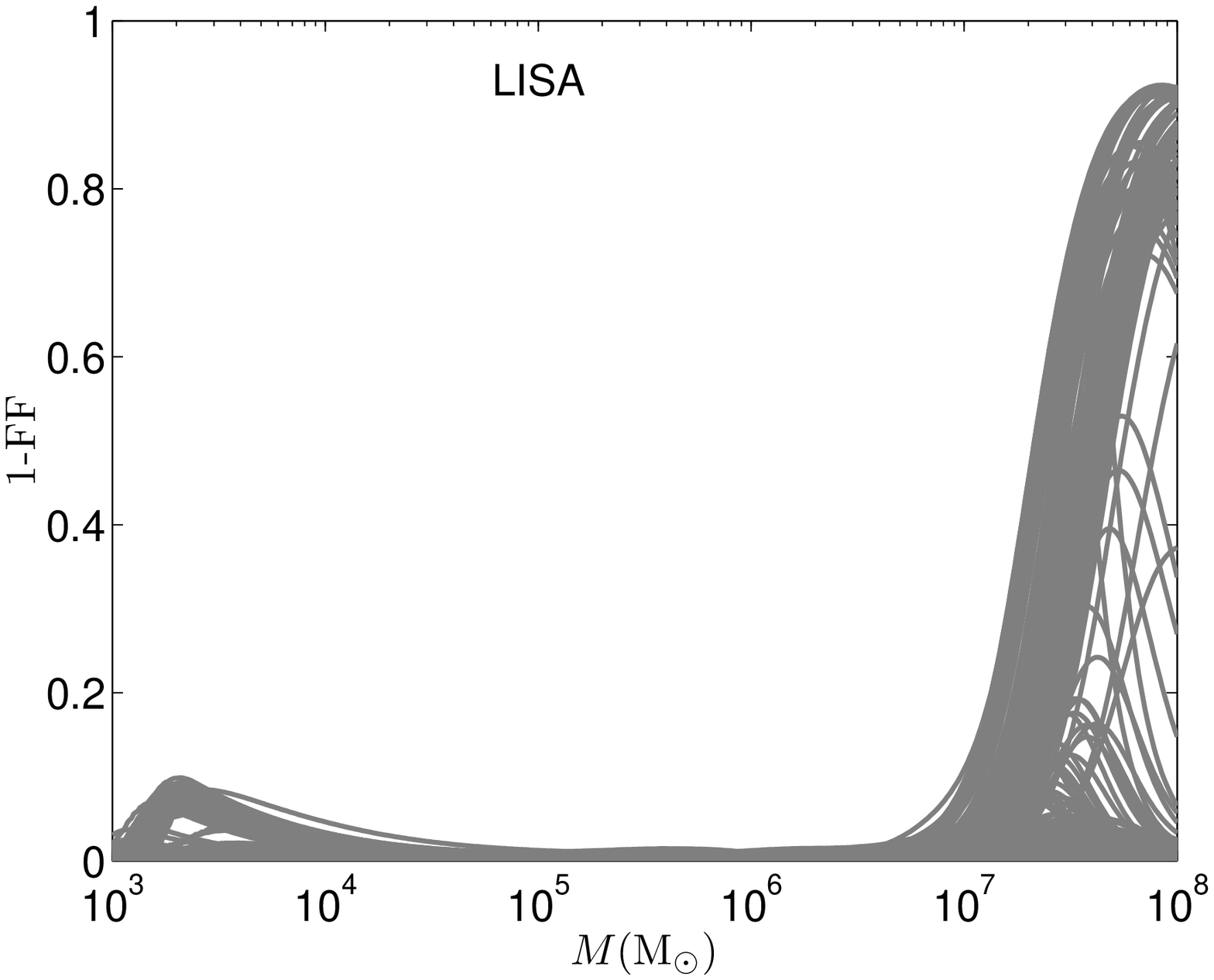}\\
\includegraphics[width=0.5\textwidth]{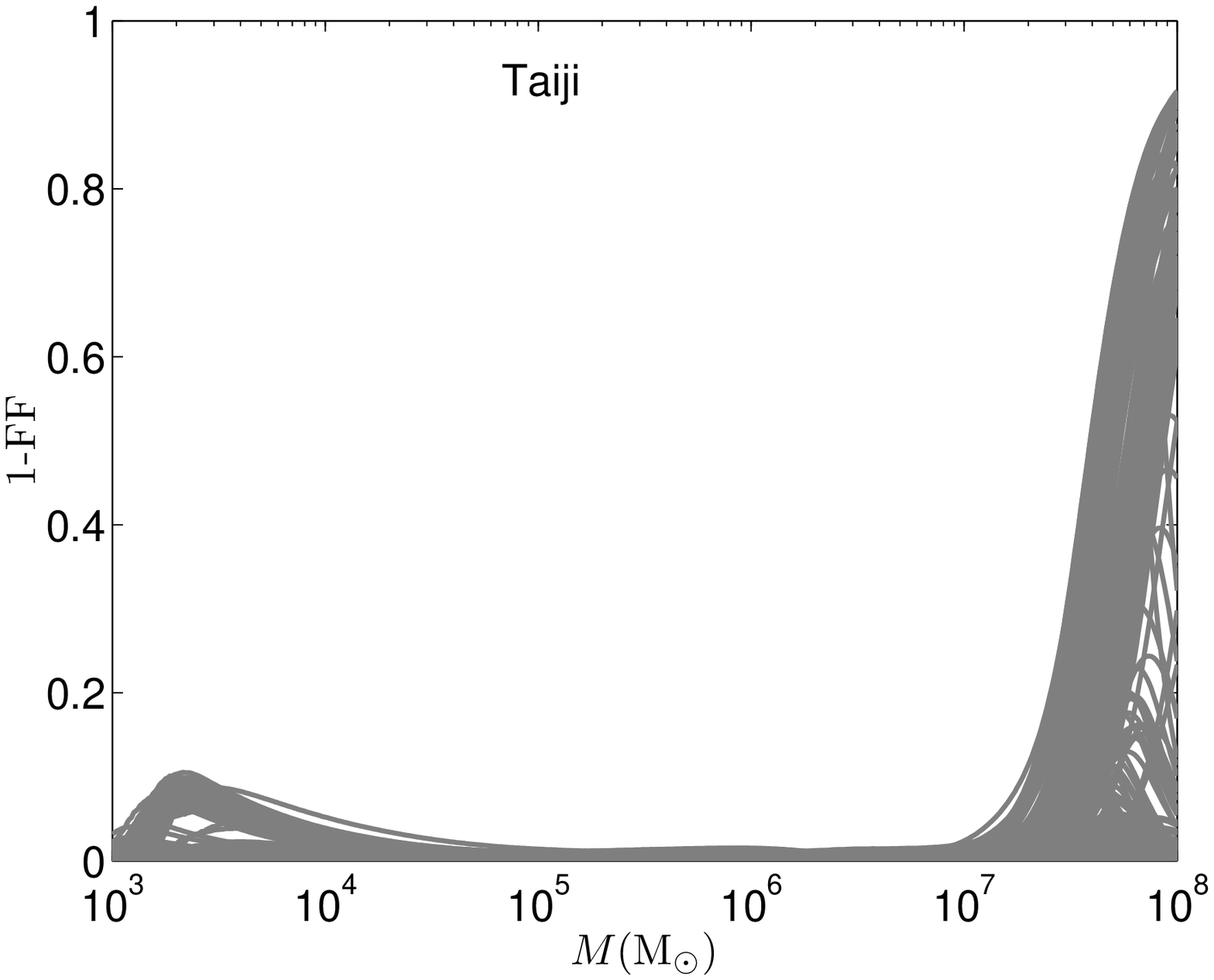}\\
\includegraphics[width=0.5\textwidth]{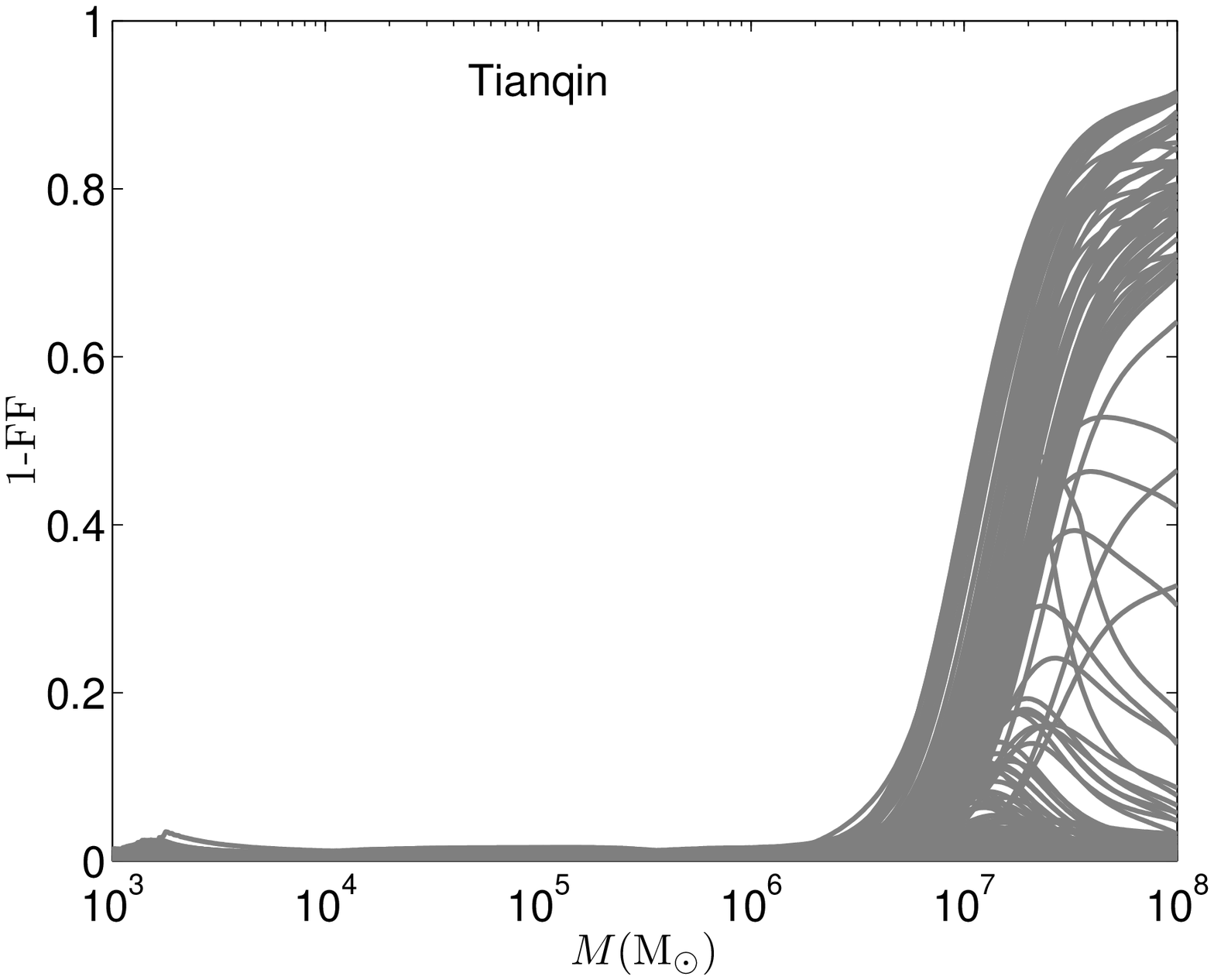}
\end{tabular}
\caption{Fitting factors between SEOBNRv4 waveforms and NR waveforms for cases listed in Table.~\ref{table1}. Three panels correspond to LISA, Taiji and Tianqin respectively.}\label{fig8}
\end{figure}

\begin{figure}
\begin{tabular}{c}
\includegraphics[width=0.5\textwidth]{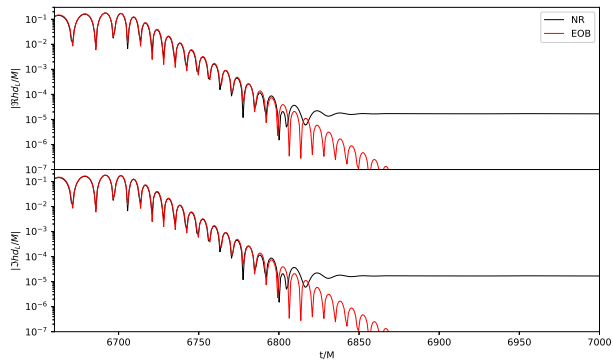}
\end{tabular}
\caption{Ringdown stage quasi-circular waveforms comparison for SEOBNRv4 (marked with `EOB') against numerical relativity (marked with `NR'). Here spinless BBH with mass ratio 6 is considered.}\label{fig9}
\end{figure}

\begin{figure}
\begin{tabular}{c}
\includegraphics[width=0.5\textwidth]{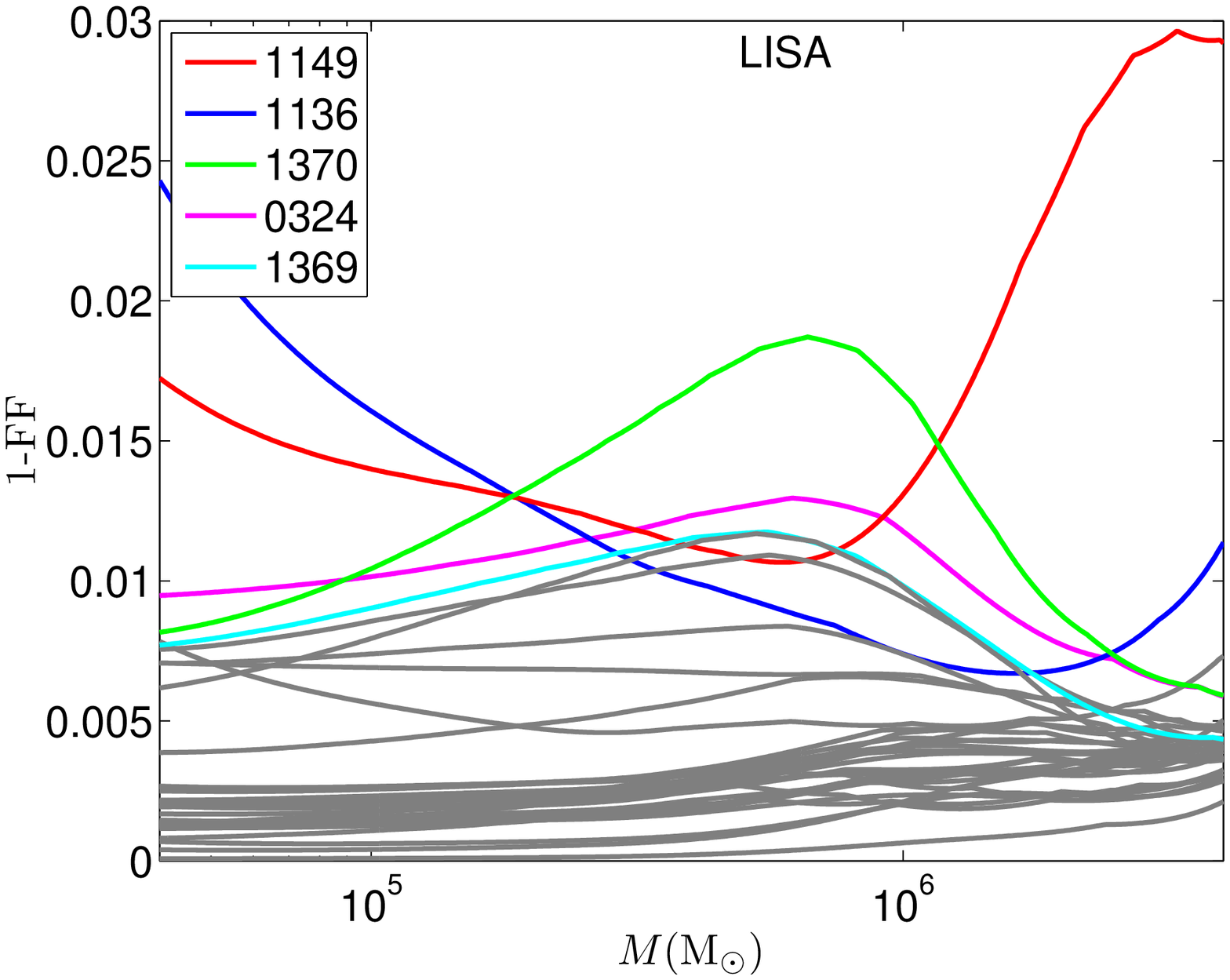}\\
\includegraphics[width=0.5\textwidth]{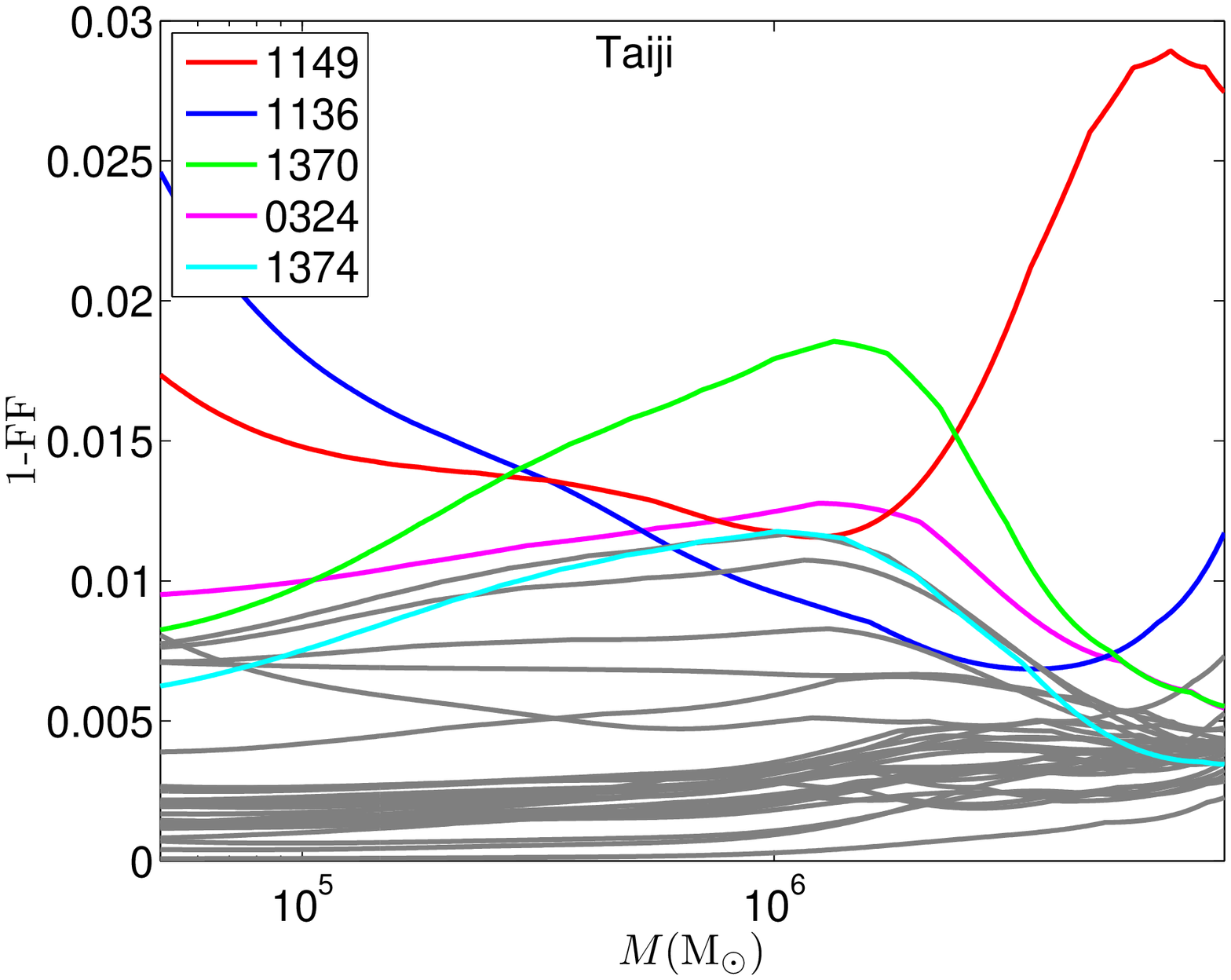}\\
\includegraphics[width=0.5\textwidth]{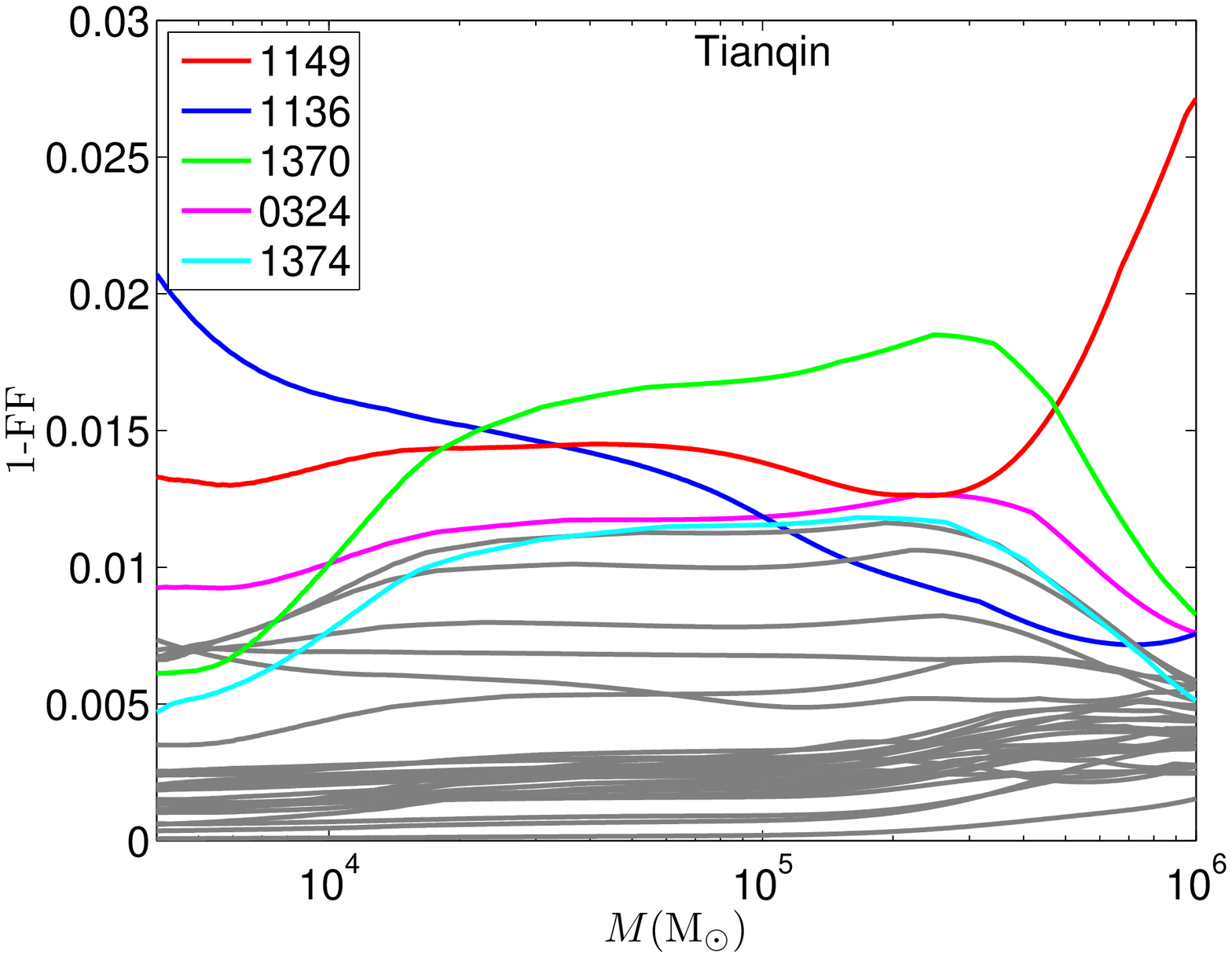}
\end{tabular}
\caption{Fitting factors between SEOBNREv4 waveforms and NR waveforms for cases listed in Table.~\ref{table2}. Three panels correspond to LISA, Taiji and Tianqin respectively.}\label{fig10}
\end{figure}

The mass range of supermassive black hole binaries for space based gravitational wave detectors is $(10^3,10^8)$M${}_\odot$. The fitting factors between SEOBNRv4 and SXS simulations listed in the Table.~\ref{table1} respect to LISA, Taiji and Tianqin respectively are plotted in the Fig.~\ref{fig8}. We can see the fitting factors are very bad for BBH with total mass larger than $10^7$M${}_\odot$. This is because when the total mass is larger than $10^7$M${}_\odot$, the later ringdown stage waveform dominates. For the later ringdown stage, both EOBNR model and NR results admit problems. Regarding EOBNR, due to the rundoff error during the calculation, the too small ringdown waveform will be donimated by numerical errors. For NR  more complicated numerical errors ruin the late ringdown stage waveform. We plot the related behavior in Fig.~\ref{fig9}. If ones restrict the mass range respectively in $(4\times10^4,4\times10^6)$M${}_\odot$, $(5\times10^4,9\times10^6)$M${}_\odot$ and $(4\times10^3,1\times10^6)$M${}_\odot$ for LISA, Taiji and Tianqin, the fitting factors can be larger than 98\%. Correspondingly we check SEOBNREv4 also for the mass range $(4\times10^4,4\times10^6)$M${}_\odot$. In Fig.~\ref{fig10} we find that the consistence between SEOBNREv4 waveforms and NR ones are all better than 97\%. For most cases the fitting factor is bigger than 99\%. 
\section{Summary and conclusion}\label{secIV}
The gravitational wave detection of compact binary objects mergers with the advanced LIGO and advanced VIRGO detectors is now a common occurrence. Although no orbit eccentricity has been found yet in the observed binary merger events \cite{Romero-Shaw:2019itr,2020MNRAS.495..466W,Nitz_2020,lenon2020measuring}, binary black hole (BBH) may form in dense stellar environments, which are expected to enter the frequency band of ground-based GW detectors with non-negligible eccentricity \cite{tagawa2020formation}. Such possibility inspires the study and modeling of eccentric BBH systems in recent years. This is because accurate waveform model can enhance the detection of eccentric BBH.

In \cite{PhysRevD.96.044028,PhysRevD.101.044049} we have constructed an accurate waveform model for mildly spin aligned binary black hole along eccentric orbit, SEOBNRE. Our SEOBNRE waveform model can accurately recover numerical relativity waveform when black hole's spin is less than 0.6. In the current paper, we upgrade previous SEOBNRE model to let it consistent to the most widely used EOBNR waveform SEOBNRv4. Correspondingly we call our previous SEOBNRE model SEOBNREv1 and the newly upgraded one SEOBNREv4. Compared to SEOBNREv1, the new model SEOBNREv4 is more robust to generate waveform, and is more accurate than SEOBNREv1. The fitting factor improves from 94.8\% to 98.2\% (c.f. the Fig.~\ref{fig5}). Most importantly, the newly upgraded SEOBNREv4 can work for highly spinning binary black holes. For example, SEOBNREv1 does not work for binary black hole with individual spin $0.7313$ and $-0.85$, while SEOBNREv4 can reach fitting factor 99.9\% to the numerical relativity waveform (c.f. the Table.~\ref{table1} and the Fig.~\ref{fig3}).

In summary, our newly upgraded waveform model SEOBNREv4 is an accurate waveform model for highly spinning, nonprecession binary black holes along eccentric orbit. The tested parameter range includes $1\leq q\lesssim10$, $-0.9\lesssim\chi_{1,2}\lesssim0.995$ and $0\leq e_0\lesssim0.6$ at reference frequency $Mf_0=0.002$ (equivalently $f_0\approx 40\frac{10{\rm M}_\odot}{M}$Hz).

Hopefully our SEOBNREv4 model may help people to find eccentric signals in the existing and future LIGO/VIRGO/KAGRA data. Such detection will not only help people to understand more the binary black hole formation mechanism \cite{2020arXiv200106492R,piran2020the,veske2020have,yu2020spin} but also assist people to test gravity theory \cite{PhysRevD.100.124032,Bing:2019,Xiaokai:2019}.

In the future it is interesting to consider tidal correction of SEOBNREv4 waveform model like \cite{PhysRevD.95.104036,PhysRevD.99.044051,PhysRevD.99.044007}. Such extension may be applied to the eccentricity estimation for GW170817 and GW190425 like events \cite{lenon2020measuring}. Relating to space based gravitational wave detector \cite{Sesana_2010}, it is not only interesting but also important to extend our SEOBNREv4 to general mass ratio cases. If ones can use the same waveform model to cover both almost equal mass binary black holes and extreme mass ratio binary black hole, which is essentially the start point of effective one body theory \cite{PhysRevD.76.064028,PhysRevD.81.084056,PhysRevD.84.044014,PhysRevD.85.024046,PhysRevLett.104.091102}, we can believe such model also valid for intermediate mass ratio binary black hole systems \cite{Mandel_2008,Mandel_2009,Han_2017,PhysRevD.84.124006,2018MNRAS.480.5160F,PhysRevD.98.063018,PhysRevD.101.081502,2020arXiv200612036V,2020arXiv200713746A}. Consequently such waveform model will be robust for space based gravitational wave detector.
\section*{Acknowledgments}
This work was supported in part by the National Key Research and Development Program of China Grant No. 2021YFC2203001 and in part by the NSFC (No.~11920101003 and No.~12021003). Z. Cao was supported by CAS Project for Young Scientists in Basic Research YSBR-006 and by ``the Interdiscipline Research Funds of Beijing Normal University".
\bibliography{refs}

\end{document}